\documentclass[3p,number,sort&compress,fleqn]{elsarticle}
\usepackage{xcolor}
\usepackage[pagebackref=false,colorlinks,linkbordercolor=white,linkcolor=cyan,citecolor=blue]{hyperref} 
\usepackage{amssymb,amsmath,latexsym}
\usepackage{bm}
\usepackage{pifont}   
\usepackage{natbib}   
\usepackage{geometry} 
\usepackage{graphicx} 
\usepackage{epstopdf} 

\newcommand*{\doi}[1]{\href{http://dx.doi.org/#1}{doi:#1}}

\newcommand{\av}[1]{\left\langle #1\right\rangle}  
\renewcommand{\atop}[2]{\genfrac{}{}{0pt}{}{#1}{#2}}
\newcommand{\sign}{\mathop{\text{sign}}} 
\renewcommand{\Im}{\mathop{\mathrm{Im}}} 
\renewcommand{\Re}{\mathop{\mathrm{Re}}} 
\newcommand{\bnabla}{\bm{\nabla}}        

\newcommand{\norm}[1]{\left|\!\left|{#1}\right|\!\right|}
\newcommand{\defi}{{:=}}                            
\newcommand{\dd}{\mathrm{d}}                        
\newcommand{\ii}{\mathrm{i}}                        
\newcommand{\ee}{\mathrm{e}}                        
\newcommand{\Pf}{\mathop{\mathrm{Pf}}}              
\newcommand{\pv}{\mathop{\mathrm{p.v.}}}            
\newcommand{\moy}[1]{\left\langle{#1}\right\rangle} 
\newcommand{\bxi}{\bm{\xi}}              
\newcommand{\bzeta}{\bm{\zeta}}          
\newcommand{\ba}{\mathbf{a}}             
\newcommand{\bb}{\mathbf{b}}             
\newcommand{\bk}{\mathbf{k}}             
\newcommand{\bl}{\mathbf{l}}             
\newcommand{\bbm}{\mathbf{m}}            
\newcommand{\bn}{\mathbf{n}}             
\newcommand{\br}{\mathbf{r}}             
\newcommand{\bs}{\mathbf{s}}             
\newcommand{\bv}{\mathbf{v}}             
\newcommand{\bx}{\mathbf{x}}             
\newcommand{\bA}{\mathbf{A}}             
\newcommand{\bL}{\mathbf{L}}             
\newcommand{\hb}{\widehat{b}}  
\newcommand{\hk}{\widehat{k}}  
\newcommand{\bhe}{\mathbf{\widehat{e}}}  
\newcommand{\bhk}{\mathbf{\widehat{k}}}  
\newcommand{\bhr}{\mathbf{\widehat{r}}}  
\newcommand{\sfC}{\mathsf{C}}             
\newcommand{\sfD}{\mathsf{D}}             
\newcommand{\sfF}{\mathsf{F}}             
\newcommand{\sfG}{\mathsf{G}}             
\newcommand{\sfI}{\mathsf{I}}             
\newcommand{\sfM}{\mathsf{M}}             
\newcommand{\sfN}{\mathsf{N}}             
\newcommand{\sfP}{\mathsf{P}}             
\newcommand{\cL}{c_{\text{L}}}

\newcommand{\cS}{c_{\text{S}}}

\newcommand{\avphi}[1]{\av{#1}_{\!\!\phi}}
\newcommand{\sGp}{\mathsf{G}^+}      
\newcommand{\sGm}{\mathsf{G}^-}      
\newcommand{\sGz}{\mathsf{G}^0}      
\newcommand{\dps}[1]{{\displaystyle #1}} 
\begin{document}
\begin{frontmatter}
\title{Causal Stroh formalism for uniformly-moving dislocations in anisotropic media: Somigliana dislocations and Mach cones}

\author[dif]{Yves-Patrick Pellegrini}
 \ead{yves-patrick.pellegrini@cea.fr}

\address[dif]{CEA, DAM, DIF, F-91297 Arpajon, France.}

*
\begin{abstract}
In this work, Stroh's formalism is endowed with causal properties on the basis of an analysis of the radiation condition in the Green tensor of the elastodynamic wave equation. The modified formalism is applied to dislocations moving uniformly in an anisotropic medium. In practice, accounting for causality amounts to a simple analytic continuation procedure whereby to the dislocation velocity is added an infinitesimal positive imaginary part. This device allows for a straightforward computation of velocity-dependent field expressions that are valid whatever the dislocation velocity ---including supersonic regimes--- without needing to consider subsonic and supersonic cases separately. As an illustration, the distortion field of a Somigliana dislocation of the Peierls-Nabarro-Eshelby-type with finite-width core is computed analytically, starting from the Green tensor of elastodynamics. To obtain the result in the form of a single compact expression, use of the modified Stroh formalism requires splitting the Green function into its reactive and radiative parts. In supersonic regimes, the solution obtained displays Mach cones, which are supported by Dirac measures in the Volterra limit. From these results, an explanation of Payton's `backward' Mach cones [R.\ G.\ Payton, Z.\ Angew.\ Math.\ Phys. \textbf{46}, 282--288 (1995)] is given in terms of slowness surfaces, and a simple criterion for their existence is derived. The findings are illustrated by full-field calculations from analytical formulas for a dislocation of finite width in iron, and by Huygens-type geometric constructions of Mach cones from ray surfaces.
\end{abstract}
\end{frontmatter}
\section{Introduction}
The computation of displacement or stress fields of dislocations in uniform motion \cite{WEER80,HIRT82,MURA87} has attracted a lot of attention in the past since the pioneering works of Frank \cite{FRAN49}, Eshelby \cite{ESHE49a}, Bullough and Bilby \cite{BULL54}, and Weertman \cite{WEER61}. The theoretical possibility of supersonic dislocations \cite{WEER80} was put forward very early \cite{ESHE56,ANG58,WEER67,WEER69b}, but had to wait until atomistic simulations to earn some credence \cite{GUMB99}. Although experimental evidence is still lacking, the quest for supersonic dislocations in metals has since gained impetus with the help of current experimental and simulation tools \cite{RUES15,HAHN16}. However, experimental evidence for supersonic dislocations is already available in systems other than metals, e.g., in dust plasma crystals \cite{NOSE11} or in relation with seismological events \cite{VALL12}.

We mainly deal hereafter with two classes of rectilinear dislocations \cite{HIRT82}. The first one is that of elementary `point-like' Volterra dislocations. The other one is that of So\-mi\-glia\-na dislocations \cite{TEOD82,LAZA16b} (sometimes called \emph{smeared-out} dislocations). Hereafter we take them with a flat core of finite width, of the specific functional kind that solves both the Peierls-Nabarro equation \cite{PEIE40,NABA47,NABA97a,SCHO05} and the Weertman equation \cite{WEER69a,WEER69b,WEER80,ROSA01} with the Frenkel (sine) pull-back force \cite{HIRT82,ROSA01}. Eshelby \cite{ESHE49a} first used this particular dislocation model to study the width of a moving dislocation. Afterwards, it was employed in further studies on motion \cite{BULL54,MARK01c,ROSA01,PELL14}. For conciseness, and for lack of a definite name, such a dislocation will tentatively be called hereafter an \emph{Eshelby dislocation}. A precise definition of it is given in Sec.\ \ref{sec:eshdis} below. Results for Volterra dislocations can be obtained as limits of results for Somgliana dislocations by letting the core width go to zero \cite{MARK01c,PELL11}. Such limits define in general distributions or pseudo-functions, and this aspect of Volterra dislocations has recently been emphasized as it is particularly important in dynamics \cite{PELL12,PELL14,PELL15,LAZA16}. More classically, it has long been known \cite{ANG58} that Volterra supersonic expressions involve Mach cones in the form of Dirac measures along cone generators \cite{CALL80,PAYT95,PELL15}. Because Dirac measures (or pseudo-functions, for that matter) cannot be plotted in any meaningful way, numerical field calculations in the supersonic regime require considering Somigliana (or other kinds of smeared-out dislocations; e.g., \cite{PELL15,LAZA16}) instead of Volterra dislocations. Somigliana dislocations can be obtained by convolution of the Volterra distributional solution with the appropriate dislocation density function and result in more regular fields \cite{ESHE49a,ESHE49b,KROU95}. As pointed out by Eshelby \cite{ESHE56} and Weertman \cite{WEER69b} in the context of isotropic elasticity, Mach cones generated by such smeared-out dislocations are spread over a finite width of order the dislocation width; thus, they can be evaluated numerically \cite{MARK01c,PELL11}.

Our emphasis is on dislocation motion in anisotropic media. An anisotropic medium sustains three wave vector-dependent bulk wave\-speeds \cite{AULD73}. Relatively to the glide direction, those wavespeeds define in non-degenerate cases three particular limiting velocities to be compared to the dislocation velocity $v$, hereafter denoted as $c_{\bf l}\leq c_{\bf i}\leq c_{\bf u}$ where the subscripts stand for ``lower'', ``intermediate'' and ``upper'', which can be computed from sections of slowness surfaces transverse to the dislocation line, as explained by Lothe in Refs.\ \cite{LOTH92,LOTH92e}.\footnote{In Ref.\ \cite{LOTH92}, $c_{\bf l}$, $c_{\bf i}$ and  $c_{\bf u}$ are denoted, respectively, $\widehat{v}$ (also $v_{\rm L}$ in \cite{LOTH92e,BARN73a}), $v'$, and $v''$.} Following Lothe's terminology (also, \cite{BARN00}), the inequality $|v|<c_{\bf l}$ defines the \emph{subsonic} range; there are three isolated \emph{transonic} velocities $v=c_{\rm l,i,u}$, and the \emph{fully supersonic} range is $|v|>c_{\rm u}$. The qualificative \emph{supersonic} applies to all velocities $|v|>c_{\bf l}$.  We shall call \emph{intersonic} the intermediate supersonic ranges $c_{\bf l}< |v|< c_{\bf i}$ and $c_{\bf i}< |v|< c_{\bf u}$.\footnote{Unfortunately, the terminology for isotropic media \cite{ANG58,WEER67} is slightly different: the range $\cS< |v|<\cL$ is often termed \emph{transonic} (but also \emph{intersonic}) and the range $|v|>\cL$  is termed \emph{supersonic}, where $\cS=c_{\rm l}=c_{\rm i}$ and $\cL=c_{\rm u}$ are the shear and dilatational wavespeeds, respectively.}

As is well-known, Stroh \cite{STRO62} devised a powerful method to deal with subsonic Volterra dislocations of arbitrary character in an anisotropic medium (see \cite{HIRT82,LOTH92b,LOTH92,BACO79} for reviews). The method requires solving a $6\times6$ eigenvalue problem ---an inexpensive task with today's computers. Moreover, the current formulation of the Stroh theory provides basic insight on anisotropic supersonic solutions \cite{STRO62,MALE70d}. For instance, the limiting velocities can be determined, as well as the opening angle of the Mach cones. Up to three Mach cones can arise (in the fully supersonic regime), involving two waves each \cite{MALE70d}. To our knowledge however, a general formula for the intensity of the Mach cones (the prefactor of the Dirac measures, for Volterra dislocations) has not been derived. Besides, no anisotropic supersonic Somigliana solution is available. This prompts us to seek general analytical means to compute supersonic dislocation fields in anisotropic media.

On another issue, the geometric configuration of Mach cones in anisotropic media can prove surprising. Indeed, Payton \cite{PAYT95} reported from a particular solution for the fields of a Volterra dislocation in a transversely anisotropic medium, the possible existence of `backwards' Mach cones, i.e., V-shaped waves with cone aperture directed \emph{towards} the direction of motion. No explanation for this counter-intuitive phenomenon has been given so far in terms of general principles. Having derived his solution for a uniformly-moving dislocation as the long-time steady-state asymptotics of a transient causal field solution, Payton emphasized the necessity of accounting for causality in field expressions to investigate Mach cones. This stems from the fact that Mach cones are built over time as caustics of expanding wavefront sets, which is a causal process. However, the classical exposition of Stroh's formalism rests on postulating \textit{a priori} functional expressions for the fields, independently of any consideration of causality \cite{STRO62,BACO79,HIRT82}. But without causality, a-causal spurious wavepackets solutions of the field equations could arise, which must be prevented.

Accordingly, the purpose of this paper is: first, to endow the Stroh formalism with causal properties, which is done in Section \ref{sec:green} starting from the Green function of the wave equation; second, to derive from this modified formalism field solutions relative to Somigliana dislocations (Section \ref{sec:eshelby}). The Volterra limit of vanishing core size is also examined, revealing the features of Mach cones (geometry, and intensity), and a simple existence criterion for Payton's `backward' cones is derived (Section \ref{sec:machcones}). In addition, full-field calculations are compared with the straightforward Huygens construction of Mach cones from ray surfaces to validate our analytical method of handling fields in the supersonic regime. A concluding discussion closes the paper (Section \ref{sec:concl}). A few technical calculations are developed in \ref{sec:distr} and \ref{sec:appflat}.

Our conventions for the Fourier transform of a space- and time-dependent function $f(\bx,t)$ are as follows:
\begin{align}
f(\bk,\omega)=\int\dd^d\!x\,\dd t\,f(\bx,t)\ee^{-\ii(\bk\cdot\bx-\omega t)},\qquad
f(\bx,t)=\int\frac{\dd^d\!k}{(2\pi)^d}\frac{\dd \omega}{2\pi}\,f(\bk,\omega)\ee^{\ii(\bk\cdot\bx-\omega t)},
\end{align}
where unless otherwise stated integrations with respect to the space variable $\bx$ and the wave vector $\bk$ are over the $d$-dimensional space $\mathbb{R}^d$ ($d=2$ or $3$), and integrations with respect to the time $t$ and the angular frequency $\omega$ are over $\mathbb{R}$. Bold and sans-serif typefaces are used, respectively, to denote vectors ($\ba$) and dyadic tensors ($\mathsf{A}$) of components $A_{ij}$. A `blackboard' typeface is used to denote fourth-order tensors ($\mathbb{A}$) of components $A_{ijkl}$.

\section{Elastodynamic Green's functions in anisotropic media}
\label{sec:green}
This section briefly reviews causality issues for the Green function (e.g.~\cite{WANG94b}, and references therein) of the anisotropic wave equation, a topic previously addressed by Budreck \cite{BUDR93}. A means of accounting for causality in the Stroh representation of the Green function for uniformly-moving sources is proposed.

\subsection{Green's functions and causality}
The tensor Green functions $\sfG(\bx,t)$ associated with the material displacement in the homogeneous medium are defined as solutions of the inhomogeneous wave equation with unit impulsive body force
\begin{align}
\label{eq:gdefeqxt}
M_{ik}G_{kj}(\bx,t)=-\delta_{ij}\,\delta(\bx)\delta(t),
\end{align}
where $\sfM$ is the wave operator
\begin{align}
M_{ij}\defi c_{ikjl}\,\partial_k\partial_l-\delta_{ij}\rho\,\partial^2_t,
\end{align}
defined in terms of the material density $\rho$, and of the anisotropic elasticity tensor $\mathbb{C}$ of components $c_{ijkl}=c_{klij}=c_{jikl}$. In the Fourier representation with wave vector $\bk$ and angular frequency $\omega$, equation (\ref{eq:gdefeqxt}) reads
\begin{align}
\left(c_{iklm}k_k k_l-\rho\,\omega^2\delta_{im}\right)G_{mj}(\bk,\omega)=\delta_{ij}.
\end{align}
Different Green functions are distinguished by boundary conditions at infinity, and are built on the following template where we introduce the acoustic tensor $N_{ij}(\bk)\defi k_k c_{iklj}k_l$ and the unit $3\times3$ matrix $\sfI$:
\begin{eqnarray}
\label{eq:fgdef}
\sfG(\bk,\omega)=[\sfN(\bk)-\rho \omega^2\sfI\,]^{-1}.
\end{eqnarray}
Its poles are solutions of the dispersion equation $\Omega(\bk,\omega)=0$, where
\begin{align}
\label{eq:omegadef}
\Omega(\bk,\omega)\defi\det\left[\sfN(\bk)-\rho \omega^2\sfI\right].
\end{align}
The physical solutions stem from prescribing a way to shift the poles off the real $\omega$-axis \cite{MORS53,BART89}.  The retarded Green function of the wave equation (denoted with a plus superscript), and the advanced one (denoted with a minus superscript) are defined as
\begin{eqnarray}
\label{eq:gdef}
\sfG^{\pm}(\bk,\omega)\defi\lim_{\eta\to 0^+}\sfG(\bk,\omega\pm\ii\eta)=\sfG(\bk,\omega\pm\ii 0^+),
\end{eqnarray}
where the last equality introduces the notation $0^+$. The positive quantity $\eta$ represents the reciprocal of some attenuation time of the waves. In lossless media $\eta$ is infinitesimal as in Eq.\ (\ref{eq:gdef}), which implements the outgoing (respectively, incoming) radiation wave condition in $\sGp$ (resp., $\sGm$). The advanced Green function $\sGm$ is useful in the computation of wave intensities and energies. The following identities hold:
\begin{align}
\label{eq:identg}
\sfG^{\pm}(-\bk,\omega)=\sfG^{\pm}(\bk,\omega),\qquad \sfG^{\pm}(\bk,-\omega)=\sfG^{\mp}(\bk,\omega).
\end{align}
Average ($\sfG^0$) and radiation ($\sfD$) parts of the retarded and advanced Green functions are introduced as
\begin{subequations}
\label{eq:statrad}
\begin{align}
\label{eq:gstatdef}
\sfG^0&\defi(\sGp+\sGm)/2,\\
\label{eq:draddef}
\sfD&\defi \sGp-\sGm,
\end{align}
\end{subequations}
so that
\begin{align}
\label{eq:gpm}
\sfG^\pm=\sfG^0\pm \sfD/2.
\end{align}
The average $\sfG^0$ is another Green function, since it evidently obeys Eq.\ (\ref{eq:gdefeqxt}). The radiation part $\sfD$ is not a Green function, but a solution of the homogeneous wave equation $M_{ik}D_{kj}(\bx,t)=0$ (i.e., a wavepacket). Definition (\ref{eq:draddef}), introduced by Dirac, is equivalent to computing $\sfD$ as the difference between outgoing and incoming fields from/on the source \cite{DIRA38}. Schwinger uses a decomposition such as (\ref{eq:gpm}) for the retarded field, and interprets $\sfG^0$ (which changes sign upon inverting the sign of time) as a reactive part which leads, upon computing powers, to inertial storage of energy in the field by the moving source; and the part $\sfD/2$, which keeps its sign under time inversion, as an irreversible radiative part leading to energy dissipation through resistive power \cite{SCHW49,ZEH92}. In relation with Green's functions, $\sfD$ is sometimes referred to as the \emph{propagator} (e.g., \cite{BART89}, where it is denoted by $K$). We use this denomination hereafter. The functions $\sGp$ and $\sfD$ are simply related through the relationship \cite{BART89,TEWA95}
\begin{subequations}
\label{eq:connectall}
\begin{align}
\label{eq:connect}
\sGp(\bx,t)&=\theta(t)\sfD(\bx,t)
\end{align}
which, combined with Eqs.\ (\ref{eq:statrad}), entails the subsidiary relations
\begin{align}
\label{eq:connect2}
\sGm(\bx,t)&=-\theta(-t)\sfD(\bx,t),\\
\sfG^0(\bx,t)&=\frac{1}{2}\text{sign}(t)\sfD(\bx,t).
\end{align}
\end{subequations}
Equation (\ref{eq:connect}) is an instance of Duhamel's principle of building solutions of an inhomogeneous problem from solutions of an homogeneous initial-value (i.e., Cauchy) one. The function $\sfD$ is subjected to initial conditions
\begin{align}
\label{eq:init}
\sfD(\bx,0)&=\mathsf{0},\qquad \partial_t\sfD(\bx,0)=\rho^{-1}\mathsf{I}\,\delta(\bx).
\end{align}
By substitution, one verifies using (\ref{eq:init}) that $\sfG^{0,\pm}$ in Eqs.\ (\ref{eq:connectall}) obey the inhomogeneous wave equation (\ref{eq:gdefeqxt}). However, whereas $\sfG^0$ is non-causal (non-physical), $\sfD$ is a physically admissible freely-moving wave packet, of a form determined by conditions (\ref{eq:init}). The causal function $\sGp$ is retrieved either by combining them according to (\ref{eq:gpm}), or by restoring causality in a multiplicative way by means of Duhamel's principle (\ref{eq:connect}).

\subsection{Eigenmode expansion of the Green tensor in anisotropic media and distributional expressions}
The Green functions $\sfG^\pm$ and $\sfG^0$, as well as $\sfD$, are distributions \cite{GELF64,KANW04}, whose explicit expressions for an anisotropic homogeneous medium are now reviewed. We write $\mathsf{N}(\bk)=\rho\,k^2 \sfC(\bhk)$, where $\bhk=\bk/k$ is the unit director of $\bk$, and $\sfC$ is the acoustic operator of components
\begin{align}
C_{ij}&=\frac{1}{\rho}\hk_k c_{iklj}\hk_l.
\end{align}
It is symmetric and admits the diagonalization
\begin{equation}
\label{eq:diag}
\sfC(\bhk)=\sum_{\alpha=1}^3 c_\alpha^2(\bhk)\,\sfP^\alpha(\bhk),
\end{equation}
where the three real and positive eigenvalues are squares of (phase) wavespeeds $c_\alpha(\bhk)>0$ for plane elastic waves that propagate in direction $\bhk$. The projectors $\sfP^\alpha(\bhk)$ are built from associated polarization eigenvectors $\bhe^\alpha(\bhk)$ that form a complete basis. Thus,
\begin{align}
\label{eq:cbas}
\sfP^\alpha(\bhk)\defi\bhe^\alpha(\bhk)\otimes\bhe^\alpha(\bhk),\qquad
\sum_{\alpha=1}^3 \sfP^\alpha(\bhk)=\mathsf{I}.
\end{align}
Definitions (\ref{eq:fgdef}) and (\ref{eq:gdef}) entail
\begin{equation}
\mathsf{G}^{\pm}(\bk,\omega)=\frac{1}{\rho}\sum_{\alpha=1}^3 \frac{\sfP^\alpha(\bhk)}{c_\alpha^2(\bhk) k^2-(\omega\pm\ii 0^+)^2}.
\end{equation}
Henceforth, and unless otherwise stated, we drop for brevity the dependence on $\bhk$ of $c_\alpha(\bhk)$ and $\sfP^\alpha(\bhk)$. Since $(\omega\pm\ii\,0^+)^2=\omega^2\pm\ii\sign(\omega)0^+$, using the Sokhotski-Plemelj formula (\ref{eq:sk}) yields  Eq.\ (\ref{eq:gdef}) in distributional form as
\begin{align}
\label{eq:gfourdistr}
\mathsf{G}^{\pm}(\bk,\omega)
=\frac{1}{\rho}\sum_{\alpha=1}^3\sfP^\alpha\left[\pv\frac{1}{c_\alpha^2 k^2-\omega^2}
\pm \ii\pi\sign\omega\,\delta\left(c_\alpha^2 k^2-\omega^2\right)\right],
\end{align}
where `$\pv$' is a principal-value prescription and $\delta$ is the Dirac distribution. Substituting (\ref{eq:gfourdistr}) into definitions (\ref{eq:statrad}), one deduces the Fourier forms of $G^0$ and $D$ as
\begin{subequations}
\begin{align}
\label{eq:gokom}
\sGz(\bk,\omega)&=\frac{1}{\rho}\sum_{\alpha=1}^3\pv\frac{\sfP^\alpha}{c_\alpha^2 k^2-\omega^2},\\
\label{eq:dkom}
\sfD(\bk,\omega)&=\frac{2\ii\pi}{\rho}\sign\omega\sum_{\alpha=1}^3\sfP^\alpha\delta\left(c_\alpha^2 k^2-\omega^2\right)=2\ii\Im\sGp(\bk,\omega).
\end{align}
\end{subequations}
Using (\ref{eq:cbas}.2) and Eq.\ (\ref{eq:decompdir}) below, Eq.\ (\ref{eq:dkom}) immediately entails
\begin{align}
\label{eq:initk}
\int\frac{\dd\omega}{2\pi}\sfD(\bk,\omega)&=\mathsf{0},\qquad\text{and}\qquad \int\frac{\dd\omega}{2\pi}(-\ii\omega)\sfD(\bk,\omega)=\frac{1}{\rho}\mathsf{I},
\end{align}
which express the initial conditions (\ref{eq:init}) in the Fourier representation.

In this paper, we consider only the two-dimensional (2D) problem, for which we now obtain $\sfD$ in space-time coordinates. Fourier inversion of $\sfD(\bk,\omega)$ is carried out by integrating over $\omega$ first. Thus, from (\ref{eq:dkom}) and the expansion
\begin{align}
\label{eq:decompdir}
\sign\omega\,\delta\left(c_\alpha^2 k^2-\omega^2\right)
&=\frac{\sign\omega}{2|\omega|}\left[\delta\left(\omega-c_\alpha k\right)+\delta\left(\omega+c_\alpha k\right)\right]
=\frac{1}{2c_\alpha k}\left[\delta\left(\omega-c_\alpha k\right)-\delta\left(\omega+c_\alpha k\right)\right],
\end{align}
we deduce
\begin{align}
\sfD(\br,t)
&=\int\frac{\dd^2\!k}{(2\pi)^2}\frac{\dd\omega}{2\pi}\sfD(\bk,\omega)\ee^{\ii(\bk\cdot\br-\omega t)}
=\frac{\ii}{8\pi^2\rho}\sum_{\alpha=1}^3\int_0^{2\pi} \dd\phi\frac{\sfP^\alpha}{c_\alpha}
\int_0^\infty \dd k\,\left[\ee^{\ii k(\bhk\cdot\br-c_\alpha t)}-\ee^{\ii k(\bhk\cdot\br+c_\alpha t)}\right],
\end{align}
where $\phi$ is the polar angle for the unit director $\bhk$. Its reference orientation for $\phi=0$ is irrelevant for the time being (a particular choice will be made in the next section). Angular integrals over the unit circle are obviously unchanged by the substitution $\bhk\to -\bhk$, which we use in the rightmost exponential, thereby transforming the difference of exponentials within brackets into $-2\ii\sin[k(c_\alpha t-\bhk\cdot\br)]$. By using next
\begin{align}
\int_0^\infty \dd k\,\sin(k x)=\pv\frac{1}{x},
\end{align}
and denoting angular averages by the following shorthand notation $\avphi{f}=(2\pi)^{-1}\int_0^{2\pi}\dd\phi\,f(\phi)$, we arrive at the integral representation
\begin{subequations}
\begin{align}
\label{eq:dint2d}
\sfD(\br,t)
=\frac{1}{(2\pi)^2\rho}\sum_{\alpha=1}^3
\avphi{\pv\frac{1}{c_\alpha(\bhk)}
\frac{\sfP^\alpha(\bhk)}{c_\alpha(\bhk) t-\bhk\cdot\br}
}.
\end{align}
Doubling the integrand and again exploiting the $\bhk\to -\bhk$  invariance of $c_\alpha(\bhk)$ and $\sfP^\alpha(\bhk)$ turns it into
\begin{align}
\label{eq:dint2dbis}
\sfD(\br,t)&=\frac{1}{2\pi\rho t}\sum_{\alpha=1}^3\avphi{\pv\frac{\sfP^\alpha(\bhk)}{c_\alpha^2(\bhk)-[\bhk\cdot(\br/t)]^2}}.
\end{align}
\end{subequations}
Combining (\ref{eq:dint2d}) and (\ref{eq:dint2dbis}), and (\ref{eq:connect}) eventually yields
\begin{align}
\label{eq:gint2d}
\sfG^+(\br,t)&=\frac{\theta(t)}{2\pi\rho}\sum_{\alpha=1}^3\avphi{\pv\frac{1}{c_\alpha(\bhk)}
\frac{\sfP^\alpha(\bhk)}{c_\alpha(\bhk) t-\bhk\cdot\br}}=\frac{\theta(t)}{2\pi\rho\,t}\sum_{\alpha=1}^3\avphi{\pv\frac{\sfP^\alpha(\bhk)}{c_\alpha^2(\bhk)-[\bhk\cdot(\br/t)]^2}}.
\end{align}
Mura provides a three-dimensional version of the intermediate expression in (\ref{eq:gint2d}), see Eq.\ (9.23) p.\ 61 in \cite{MURA87}. Furthermore, Tewary \cite{TEWA95} gives analogous representations of the 3D time-dependent Green function based on Duhamel's principle, and Radon transforms \cite{WANG94b}. However, for 2D problems Radon-transform and Fourier methods such as above are identical \cite{DEAN83,LIUL96}.

Using the mutual orthogonality of the eigenvectors, one verifies that $\sfD(\br,t)$ in the form (\ref{eq:dint2d}) solves the homogeneous wave equation $M_{ik}D_{kj}(\br,t)=0$, in which gradients must now be considered as two-dimensional with respect to $\br$. Moreover, the initial-value conditions (\ref{eq:init}) must hold with $\delta(\bx)$ replaced by the two dimensional $\delta(\br)$. To check, we note first that, obviously, $\sfD(\br,0)=\mathsf{0}$ by symmetry. Next, differentiating (\ref{eq:dint2d}) with respect to time we get (with a finite-part prescription, hereafter denoted by `$\Pf$')\footnote{
In the sense of distributions, $(\pv x^{-1})'=-\Pf x^{-2}$ \cite{KANW04}.
}
\begin{align}
\label{eq:ddint2d}
\rho\frac{\partial\sfD}{\partial t}(\br,t)&=-\frac{1}{2\pi}\sum_{\alpha=1}^3\avphi{\Pf\frac{\sfP^\alpha(\bhk)}{[c_\alpha(\bhk) t-\bhk\cdot\br]^2}}.
\end{align}
Taking $t=0$ in this expression allows one to retrieve the expected condition $\partial_t\sfD(\br,0)=\rho^{-1}\mathsf{I}\delta(\br)$, on account of the completeness relation (\ref{eq:cbas})${}_2$, and of the interesting distributional identity (\ref{sec:distr})
\begin{align}
\label{eq:curious}
\avphi{\Pf\frac{1}{(\bhk\cdot\br)^2}}&=-2\pi\delta(\br).
\end{align}

\subsection{Connection with Stroh's formalism: propagator and Green's functions in space-time representation}
\label{sec:constroh}
The integral form (\ref{eq:dint2dbis}) of $\sfD(\br,t)$ is now computed exactly by means of Stroh's formalism following Hirth and Lothe's notations and conventions \cite{HIRT82}. The formal connection between the dynamic kernels and those showing up in the problem of sources moving at constant velocity is revealed by introducing a fictitious `velocity' variable $\bv=\br/t$ \cite{WU00}, and modified elastic constants \cite{SAEN53,TEUT62a}
\begin{align}
\label{eq:modc}
\widetilde{c}_{ijkl}&=c_{ijkl}-\rho v_j v_k \delta_{il}.
\end{align}
For two real vectors $\ba$ and $\bb$ we abbreviate by $(ab)$ \cite{HIRT82} the $3\times 3$ matrix of components
\begin{align}
\label{eq:hlbrack}
(ab)_{ij}&=a_k \widetilde{c}_{iklj} b_l.
\end{align}
Due to the minor Voigt symmetry of the elasticity tensor $c_{ijkl}$, different indexing conventions for the $v^2$ term in Eq.\ (\ref{eq:modc}) are found in the literature (e.g., \cite{SAEN53}). The bracket notation (\ref{eq:hlbrack}) must be defined consistently with the chosen convention so as to reproduce the results below. From (\ref{eq:modc}), (\ref{eq:hlbrack}) and the eigenvalue decomposition (\ref{eq:diag}), (\ref{eq:cbas}) one deduces that
\begin{align}
(\hk\hk)&=\rho\left[\sfC(\bhk)-(\bhk\cdot\bv)^2\sfI\right]=\rho\sum_{\alpha=1}^3 \left[c^2_\alpha(\bhk)-(\bhk\cdot\bv)^2\right]\sfP^\alpha(\bhk),
\end{align}
which yields the inverse tensor \cite{LOTH92} (Eq.\ (194) in that reference)
\begin{align}
\label{eq:kkm1}
(\hk\hk)^{-1}=\frac{1}{\rho}\sum_{\alpha=1}^3 \frac{\sfP^\alpha(\bhk)}{c^2_\alpha(\bhk)-(\bhk\cdot\bv)^2}.
\end{align}
Substituting this expression into Eq.\ (\ref{eq:dint2dbis}) then yields the propagator in the form
\begin{align}
\label{eq:davexpr}
\sfD(\br,t)=\frac{1}{2\pi t}\pv\avphi{(\hk\hk)^{-1}}.
\end{align}
\begin{figure}[!ht]
\centering
\includegraphics[width=7cm]{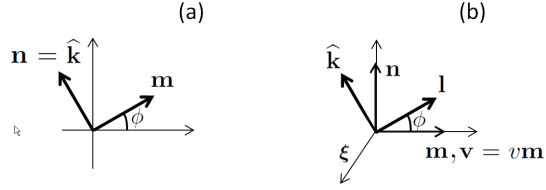}
\caption{\label{fig:fig1} Rotating bases. (a) as used in Sec.\ \ref{sec:constroh}. (b) as used in Sec.\ \ref{sec:eshdis}.}
\end{figure}

This expression can be evaluated by means of the Stroh formalism. Indeed, introduce the unit vector $\bbm=(\cos\phi,\sin\phi)$ and identify $\bhk$ with the complementary orthogonal vector $\bn=(-\sin\phi,\cos\phi)$ (Fig.\ \ref{fig:fig1}a); note that we do not assume here that $\bv\propto\bbm$. Then, one recognizes in (\ref{eq:davexpr}) one of the angular averages that enter the well-known Barnett-Lothe integral version of Stroh's theory \cite{BARN73a,HIRT82}. The latter is briefly summarized hereafter (see, e.g., \cite{HIRT82,LOTH92b} for details). It hinges on using the $6\times 6$ nonsymmetric matrix \cite{STRO62}
\begin{align}
\label{eq:ndef}
\mathcal{N}=-\left(
\begin{array}{cc}
(nn)^{-1}\cdot(nm) & (nn)^{-1}\\
(mn)\cdot(nn)^{-1}\cdot(nm)-(mm)& (mn)\cdot(nn)^{-1}\\
\end{array}
\right).
\end{align}
The associated eigenvalue problem
\begin{align}
\label{eq:eigsyst}
\mathcal{N}\cdot\bzeta=p\,\bzeta,
\end{align}
involves the right eigenvector $\bzeta=(\bA,\bL)$, which defines from its components ordered as $\bzeta=(A_1,A_2,A_3,$ $L_1,L_2,L_3)$ two 3-vectors $\bA$ and $\bL$ that correspond to polarizations of the displacement, and traction vectors in the plane $\bn\cdot\br=0$, respectively. The matrix $\mathcal{N}$ has six eigenvalues $p^\alpha$, $\alpha=1,\ldots,6$ and associated eigenvectors $\bzeta^\alpha=(\bA^\alpha,\bL^\alpha)$. These vectors are normalized such that
\begin{align}
\bA^\alpha\cdot\bL^\beta+\bA^\beta\cdot\bL^\alpha=\delta_{\alpha\beta}.
\end{align}

For $|v|<c_{\rm l}$, the eigenvalues constitute three pairs of conjugate complex numbers conventionally labelled so that $\Im p^\alpha\geq 0$ for $1\leq \alpha\leq 3$, and such that $\overline{p}^\alpha=p^{\alpha+3}$ (the overbar denotes the complex conjugate). From (\ref{eq:eigsyst}), the same property is inherited by $\bA^\alpha$ and  $\bL^\alpha$, and $\overline{\bzeta}{}^\alpha=\bzeta^{\alpha+3}$ for $1\leq \alpha\leq 3$. A key property of the formalism is that both vectors $\bA_\alpha$ and $\bL_\alpha$ are independent of the orientation angle $\phi$. However, $p^\alpha$ depends on it as
\begin{align}
\label{eq:ptan}
p^\alpha(\phi)=\tan(\psi_\alpha-\phi),
\end{align}
where $\psi_\alpha$ is a complex constant. This expression shows that $\Im\psi_\alpha$ and $\Im p^\alpha$ are of same signs \cite{CHAD77}. As $v$ increases, two $p^\alpha$s become real-valued, as well as their associated angles $\psi_\alpha$, each time $v$ crosses one of the transonic velocities. In the supersonic range all the $p^\alpha$ are real. This point is further discussed in Sec.\ \ref{sec:machcones} in connection with Mach cones.

An important quantity to be used hereafter is the angular average $\pv\avphi{p^\alpha}$, defined as a principal value.  Whenever $\psi_\alpha$ has a nonzero imaginary part, the integral over angles is nonsingular, with the result $\pv\avphi{p^\alpha}=\avphi{p^\alpha}=\pm\ii$, where the sign is that of $\Im p^\alpha$ \cite{BARN73b,CHAD77} (by contour integration on the unit circle with integration variable $z=\exp\ii\phi$). By contrast, when $\psi_\alpha$ is real the integral is divergent at $\phi=\psi_\alpha\pm\pi/2$ mod.\ $2\pi$, and the indispensable `$\pv$' prescription makes it finite by handling these singularities as principal values. Using (\ref{eq:ptan}), one easily gets from an obvious change of variables the result $\pv\avphi{p^\alpha}=\pv\avphi{\tan\phi}=0$. Thus, introducing
\begin{align}
s_\alpha&=\sign\Im p^\alpha,
\end{align}
we have
\begin{align}
\label{eq:palphaav}
\pv\avphi{p^\alpha}
=\left\{
\begin{array}{ll}
\ii\,s_\alpha & \Im p^\alpha\not= 0\\
0 & \Im p^\alpha= 0
\end{array}
\right.
.
\end{align}

Expanding the matrix equation (\ref{eq:eigsyst}) shows that the vectors $\bL^\alpha$ are connected to the $\bA^\alpha$ by the relationship
\begin{align}
\label{eq:la}
\bL^\alpha=-[(nm)+p^\alpha(nn)]\cdot\bA^\alpha,
\end{align}
and that the existence of nonzero eigenvectors $\bA^\alpha$ requires the eigenvalues $p=p^\alpha$ to be determined in terms of $v$ by the equation
\begin{align}
\label{eq:pv}
\Delta(p,v)\defi\det\left\{(mm)+p[(mn)+(nm)]+p^2(nn)\right\}&=0.
\end{align}
As a consequence, $\mathcal{N}$ can be shown to admit the decomposition
\begin{align}
\label{eq:ndecomp}
\mathcal{N}&=\sum_{\alpha=1}^6 p^\alpha \bzeta^\alpha\otimes(\mathcal{V}\cdot\bzeta^\alpha),\quad\text{where}\quad
\mathcal{V}=\left(
\begin{array}{cc}
0 & \sfI\\
\sfI & 0\\
\end{array}
\right).
\end{align}
Identifying expressions (\ref{eq:ndef}) and (\ref{eq:ndecomp}), decompositions in the form of sum rules of the block matrices that make up $\mathcal{N}$ follow; notably,
\begin{subequations}
\begin{align}
\label{eq:nnnmm1}
(nn)^{-1}\cdot(nm)&=-\sum_{\alpha=1}^6 p^\alpha(\phi)\bA^\alpha\otimes\bL^\alpha,\\
\label{eq:nnm1}
(nn)^{-1}&=-\sum_{\alpha=1}^6 p^\alpha(\phi)\bA^\alpha\otimes\bA^\alpha.
\end{align}
\end{subequations}
Moreover, the following closure relations hold
\begin{align}
\label{eq:cAA}
\sum_{\alpha=1}^6 \bA^\alpha\otimes\bA^\alpha=\sum_{\alpha=1}^6 \bL^\alpha\otimes\bL^\alpha&=\mathsf{0},\qquad
\sum_{\alpha=1}^6 \bA^\alpha\otimes\bL^\alpha=\sum_{\alpha=1}^6 \bL^\alpha\otimes\bA^\alpha=\mathsf{I}.
\end{align}

We can now compute $\sfD(\br,t)$ by substituting expression (\ref{eq:nnm1}) into Eq.\ (\ref{eq:davexpr}), which results in
\begin{align}
\sfD(\br,t)=-\frac{1}{2\pi t}\sum_{\alpha=1}^6 \pv\avphi{p^\alpha(\phi)}\bA^\alpha\otimes\bA^\alpha.
\end{align}
Emphasizing the functional dependence of the vectors, and using (\ref{eq:palphaav}) one concludes that
\begin{align}
\label{eq:dstroh}
\sfD(\br,t)=\frac{1}{\pi t}\Im\sum_{\atop{\alpha=1,2,3}{\Im p^\alpha > 0}}\bA^\alpha(\bv)\otimes\bA^\alpha(\bv)\qquad (\bv=\br/t),
\end{align}
where the sum is restricted to indices of eigenvalues with strictly positive imaginary part, and where com\-plex-con\-ju\-ga\-tion properties have been used. Apart from different sign and normalization conventions, Equ.\ (\ref{eq:dstroh}) is equivalent to Wu's equations (3.12--13) \cite{WU00}. Although expression (\ref{eq:dstroh}) is sometimes referred to as a Green's function (e.g., \cite{WU00}), the above derivation clarifies its nature as a propagator. The Green functions $\sfG^{0,\pm}(\br,t)$ stem from multiplying (\ref{eq:dstroh}) by the appropriate time-dependent prefactor, as in Eqs.\ (\ref{eq:connectall}).

\subsection{Uniformly moving sources and analytically-continued representations}
Up to now, the vector $\bv$ was a shorthand for $\br/t$. In the rest of the paper, it will denote a `true' velocity. So, the dislocation moves at \emph{constant} velocity $\bv=v\,\bbm$, where the scalar $v$ can be of any sign. The unit vector $\bn$, such that $\bbm\cdot\bn=0$ is the slip-plane normal. The dislocation line is oriented along $\bxi=\bbm\times\bn$, and the plane spanned by $\bbm$ and $\bn$ is the co-called \emph{sagittal plane}.

Then, the components of the source tensor are all of the type $f(\br,t)=f(\br-\bv t)$, of Fourier transform $f(\bk,\omega)=(2\pi)\delta(\omega-\bv\cdot\bk)f(\bk)$. The induced field components all have the following form, where the integral on $\br'$ is over all $d$-dimensional space:
\begin{align}
\phi_{ij}(\br,t)
&=\int\dd^d\!r'\int_{-\infty}^{+\infty}\dd t\,G^+_{ij}(\br-\br',t-t')f(\br'-\bv t')\nonumber\\
\label{eq:invfour}
&=\int\frac{\dd^d\!k}{(2\pi)^d}\int_{-\infty}^{+\infty}\frac{\dd \omega}{2\pi} \ee^{\ii(\bk\cdot\br-\omega t)}\,G^+_{ij}(\bk,\omega)f(\bk,\omega)
=\int\frac{\dd^d\! k}{(2\pi)^d}\ee^{\ii\bk\cdot(\br-\bv t)}\,G^+_{ij}(\bk,\bk\cdot\bv)f(\bk),
\end{align}
where the last line is a well-known representation of the field in terms of inverse Fourier transforms. In this expression, the fundamental quantity is $\sfG^+(\bk,\bk\cdot\bv)$, namely,
\begin{align}
\sfG^+(\bk,\bk\cdot\bv)
&=\frac{1}{\rho}\sum_{\alpha=1}^3 \frac{\sfP^\alpha}{k^2 c_\alpha^2-(\bk\cdot\bv+\ii 0^+)^2}
=\frac{1}{\rho}\sum_{\alpha=1}^3 \frac{\sfP^\alpha}{k^2 c_\alpha^2-(\bk\cdot\bv)^2-\ii\sign(\bhk\cdot\bv)0^+}\nonumber\\
\label{eq:gkkv}
&=\pv\frac{1}{\rho}\sum_{\alpha=1}^3 \frac{\sfP^\alpha}{k^2 c_\alpha^2-(\bk\cdot\bv)^2}+\frac{\ii\pi}{\rho}\sign(\bhk\cdot\bv)\sum_{\alpha=1}^3\sfP^\alpha\delta\left(k^2 c_\alpha^2-(\bk\cdot\bv)^2\right).
\end{align}

Hereafter, motion in anisotropic media will be addressed on the basis of formula (\ref{eq:invfour}). To this aim, we need to write $\sfG^+(\bk,\bk\cdot\bv)$ in a form computable from the Stroh formalism. This is possible only if we can account for the prescription $+\ii 0^+$ by modifying the elastic tensor \emph{in a way that does not depend on $\bhk$}; otherwise, the crucial property that $\bA^\alpha$ and $\bL^\alpha$ do not depend on $\phi$ (see previous section) would be destroyed.

A convenient way to do this ---the central idea of the paper--- consists in shifting the algebraic velocity by a small positive imaginary quantity, thus introducing the complex velocity vector
\begin{align}
\bv^\epsilon&=\bv+\ii\epsilon\bbm=(v+\ii\epsilon)\bbm,
\end{align}
where $\epsilon$ is an infinitesimal positive number of irrelevant exact magnitude. Then, assuming $\bhk\cdot\bbm\not=0$,
\begin{align}
(\bhk\cdot\bv^\epsilon)^2&\simeq (\bhk\cdot\bbm)^2(v^2+2\ii\epsilon\sign(v))\simeq(\bhk\cdot\bv)^2+\ii\epsilon\sign(v).
\end{align}
Accordingly, with $\bv$ now the true velocity, we modify S\`{a}enz's velocity-dependent `elastic constants' (\ref{eq:modc}) into the complex-valued elastic moduli
\begin{align}
\label{eq:modcmplx}
\widetilde{c}_{ijkl}^{\,\epsilon}&\defi c_{ijkl}-\rho v^\epsilon_j v^\epsilon_k\delta_{il}.
\end{align}
Further introducing $(\hk\hk)_{\epsilon,il}\defi \hk_j\widetilde{c}_{ijkl}^{\,\epsilon}\hk_k$, it follows that
\begin{align}
(\hk\hk)_{\epsilon,il}&\simeq\hk_jc_{ijkl}\hk_k-\rho[(\bhk\cdot\bv)^2+\ii\epsilon\sign(v)]\delta_{il}.
\end{align}
Upon diagonalizing this expression in the polarization basis, we obtain
\begin{align}
(\hk\hk)_\epsilon&\simeq\rho \sum_{\alpha=1}^3\sfP^\alpha[c_\alpha^2-(\bhk\cdot\bv)^2-\ii\epsilon\sign(v)].
\end{align}
Furthermore introducing the function
\begin{align}
\label{eq:fkv}
\sfF^+_\epsilon(\bk,\bv)=\frac{1}{k^2}(\hk\hk)_\epsilon^{-1},
\end{align}
il follows that
\begin{align}
\label{eq:limfkv}
\lim_{\epsilon\to 0^+}\sfF^+_\epsilon(\bk,\bv)&=
\pv\frac{1}{\rho}\sum_{\alpha=1}^3 \frac{\sfP^\alpha}{ k^2c_\alpha^2-(\bk\cdot\bv)^2}+\frac{\ii\pi}{\rho}\sign(v)\sum_{\alpha=1}^3\sfP^\alpha\delta\left(k^2 c_\alpha^2-(\bk\cdot\bv)^2\right).
\end{align}
Since $\sign(\bhk\cdot\bv)=\sign(v)\sign(\bhk\cdot\bbm)$, comparing (\ref{eq:limfkv}) with (\ref{eq:gkkv}) shows that
\begin{align}
\label{eq:gwithsign}
\sfG^+(\bk,\bk\cdot\bv)&=\lim_{\epsilon\to 0}
\left[\Re+\ii\sign(\bhk\cdot\bbm)\Im\right]\sfF^+_{\epsilon}(\bk,\bv).
\end{align}
One deduces from definitions (\ref{eq:gokom}) and (\ref{eq:dkom}) and the above that
\begin{subequations}
\label{eq:g0dkv}
\begin{align}
\label{eq:g0kv}
\sfG^{0}(\bk,\bk\cdot\bv)&=\lim_{\epsilon\to 0}\frac{1}{k^2}\Re (\hk\hk)^{-1}_{\epsilon},\\
\label{eq:dkv}
\sfD(\bk,\bk\cdot\bv)&=2\ii\sign(\bhk\cdot\bbm)\lim_{\epsilon\to 0}\frac{1}{k^2}\Im (\hk\hk)_\epsilon^{-1}.
\end{align}
\end{subequations}
As will be made clear in the following, the Stroh formalism can now be used to compute $(\hk\hk)_\epsilon^{-1}$, thanks to identity (\ref{eq:nnm1}). The Green functions $\sfG^{\pm}(\bk,\bk\cdot\bv)$ are then retrieved by combining expressions (\ref{eq:g0dkv}) according to (\ref{eq:gpm}). To summarize, the causal Green function relevant to a source in uniform motion cannot be directly obtained from the Stroh formalism because of the factor $\sign(\bhk\cdot\bbm)$ in (\ref{eq:gwithsign}). However, taken separately, its real (reactive) and imaginary (radiative) parts can be, to be reassembled afterwards to retrieve  $\sfG^{\pm}$.

Our assumption that $\smash{\bhk\cdot\bbm\not=0}$ does not impair the generality of the above derivation, since the Dirac contributions to the Green function (\ref{eq:gkkv}) vanish anyway if $\bhk\cdot\bbm=0$.

\section{Elastodynamic kernels and fields induced by a uniformly moving Eshelby dislocation}
\label{sec:eshelby}
This section is devoted to obtaining the distortion and stress fields produced by a uniformly moving Somigliana dislocation, and further specialized into a dislocation of the Eshelby type.

\subsection{Somigliana dislocations, elastodynamic kernels and fields}
The Somigliana dislocation is represented by the plastic eigenstrain tensor $\beta^{\rm p}_{ij}(\br,t)$, which constitutes the source of the elastodynamic fields in the Green's function approach \cite{MURA87}. The dislocation, moving at constant velocity $\bv$, is such that in the direct and Fourier representations (respectively),
\begin{subequations}
\label{eq:betaexpr}
\begin{align}
\label{eq:betaexpr1}
\beta_{ij}^{\rm p}(\br,t)&=\beta_{ij}^{\rm p}(\br-\bv t),\\
\label{eq:betaexpr2}
\beta_{ij}^{\rm p}(\bk,\omega)&=(2\pi)\delta(\omega-\bv\cdot\bk)\beta_{ij}^{\rm p}(\bk).
\end{align}
\end{subequations}
Its shape, assumed rigid because of uniform motion, is completely characterized by $\beta_{ij}^{\rm p}(\br)$, or by $\beta_{ij}^{\rm p}(\bk)$ in the Fourier representation. The material displacement induced by the dislocation in the surrounding medium is of the form (\ref{eq:invfour}), and reads
\begin{eqnarray}
\label{eq:uk}
u_i(\br,t)=-\ii\int \frac{\dd^2\!k}{(2\pi)^2} G^+_{ij}(\bk,\bk\cdot\bv) k_k c_{jklm}\beta^{\rm p}_{lm}(\bk)\ee^{\ii \bk\cdot(\br-\bv t)}.
\end{eqnarray}
Introducing generic response kernels $\mathbb{B}^{0,\pm}$ of components\footnote{
Another form stems from writing $\delta_{jl}=G_{jp}G^{-1}_{pl}$ in (\ref{eq:bline1}) and reorganizing terms (\cite{MURA87}, p.\ 351):
\begin{align*}
B_{ijkl}(\bk,\omega)
&=[(\delta_{in}\delta_{mk}-\delta_{ik}\delta_{nm}) k_n k_q c_{pqml}-\rho\,\omega^2\delta_{ik}\delta_{pl}]G_{jp}(\bk,\omega)
= (\epsilon_{oin}\epsilon_{omk} k_n k_q c_{pqml}-\rho\,\omega^2\delta_{ik}\delta_{pl})G_{jp}(\bk,\omega).
\end{align*}
}
\begin{subequations}
\begin{align}
\label{eq:bline1}
B^{0,\pm}_{ijkl}(\bk,\omega)&= k_i G^{0,\pm}_{jp}(\bk,\omega) k_q c_{pqkl}-\delta_{ik}\delta_{jl},
\end{align}
the elastic distortion $\beta_{ij}\defi\partial_i u_j-\beta^{\rm p}_{ij}$ reads
\begin{align}
\label{eq:betaB}
\beta_{ij}(\br,t)&=\int\frac{\dd^2\!k}{(2\pi)^2} B^+_{ijkl}(\bk,\bk\cdot\bv)\beta^{\rm p}_{kl}(\bk)\ee^{\ii \bk\cdot(\br-\bv t)}.
\end{align}
\end{subequations}
Decomposition (\ref{eq:gpm}) of the Green function into reactive and radiative parts induces a similar additive decomposition of $\mathbb{B}^+$ that carries over to the fields. Hereafter, the reactive and radiative parts of $\beta_{ij}$ are denoted with a superscript $0$ or $D$, respectively. Explicitly, we shall write
\begin{align}
\mathbb{B}^+=\mathbb{B}^0+\mathbb{B}^D,
\end{align}
where
\begin{subequations}
\label{eq:bline}
\begin{align}
\label{eq:bline0}
B^0_{ijkl}(\bk,\omega)&=k_i G^0_{jp}(\bk,\omega) k_q c_{pqkl}-\delta_{ik}\delta_{jl},\\
\label{eq:blineD}
B^D_{ijkl}(\bk,\omega)&=\frac{1}{2}k_i D_{jp}(\bk,\omega) k_q c_{pqkl},
\end{align}
\end{subequations}
from which $\beta^0_{ij}$ and $\beta^D_{ij}$ are determined by expressions akin to (\ref{eq:betaB}). The stress follows from the generalized Hooke law as $\sigma_{ij}=c_{ijkl}\beta_{kl}$.

\subsection{Eshelby dislocation}
\label{sec:eshdis}
We now specialize to the Eshelby dislocation. As recalled in the Introduction, it is a natural solution of the (static) Peierls-Nabarro and (steady-motion) Weertman models for the dislocation core shape. As it has been used as well to represent non-uniformly moving dislocations \cite{ESHE53,MARK01b,MARK01c,PELL10,PELL12,PELL14}, it is of special theoretical interest. The Eshelby dislocation has a core density function per unit Burgers vector of a simple (Lorentzian-type) analytic structure, namely
\begin{align}
\label{eq:coreflat}
\rho_a(x)=\frac{1}{\pi}\frac{a}{x^2+a^2}=\frac{1}{2\ii\pi}\left(\frac{1}{x-\ii a}-\frac{1}{x+\ii a}\right),
\end{align}
where, to avoid dragging along factors $1/2$, the quantity $a$ is introduced as the \emph{half core width} rather than as the core width. This density depends solely on the in-plane Cartesian coordinate $x=\br\cdot\bbm$ in the direction of motion. We introduce the co-moving position vector $\br'$ and abscissa $x'$ defined as
\begin{align}
\br'\equiv (x',y)=\br-\bv t.
\end{align}
The dislocation is flat with respect to the out-of-plane coordinate $y$. The functions $\rho_a$ define a $\delta$-sequence such that $\rho_0(x)=\lim_{a\to 0}\rho_a(x)=\delta(x)$. In this limit $a\to 0$, the Eshelby dislocation reduces to a Volterra dislocation. The associated plastic eigenstrain is localized on the slip plane, and reads
\begin{align}
\label{eq:disc}
\beta^{\rm p}_{ij}(\br')=n_i \hb_j\eta(x')\delta(y),\qquad\text{where}\qquad\eta(x)=\frac{b}{\pi}\left(\frac{\pi}{2}-\arctan\frac{x}{a}\right).
\end{align}
Equation (\ref{eq:disc}) is our definition of the Eshelby dislocation. The slip discontinuity $\eta(x)$ spans the whole real axis. It reduces to $n_i b_j\theta(-x')\delta(y)$ in the Volterra limit $a\to 0$. Conversely, Eq.\ (\ref{eq:disc}) is retrieved by taking the convolution of the latter Volterra expression by the density (\ref{eq:coreflat}). For further purposes, we note that
\begin{align}
\label{eq:rhoeta}
\rho_a(x)=-\frac{1}{b}\eta'(x).
\end{align}

Since in the Fourier domain $\rho_a(k_x)=\ee^{-a|k_x|}=\ee^{-a k|\bhk\cdot\bbm|}$ we deduce from the Fourier transform of $\theta(-x)$ that
\begin{align}
\label{eq:betapflat}
\beta^{\rm p}_{ij}(\bk)&=\ii n_i b_j\frac{\ee^{-a k|\bhk\cdot\bbm|}}{\bk\cdot\bbm+\ii 0^+}.
\end{align}

We begin by computing the reactive distortion $\beta^0_{ij}$ from (\ref{eq:bline0}). It reads
\begin{align}
\beta^0_{ij}(\br,t)&=\int\frac{\dd^2\!k}{(2\pi)^2} B^0_{ijkl}(\bk,\bk\cdot\bv)\beta^{\rm p}_{kl}(\bk)\ee^{\ii \bk\cdot\br'}\nonumber\\
\label{eq:beta0}
&=\ii\,b_l\int\frac{\dd^2\!k}{(2\pi)^2}\left[k_i G^0_{jp}(\bk,\bk\cdot\bv) k_q c_{pqkl}-\delta_{ik}\delta_{jl}\right]\frac{n_k \ee^{\ii \bk\cdot\br'-a k|\bhk\cdot\bbm|}}{\bk\cdot\bbm+\ii 0^+}.
\end{align}
The term within square brackets has zero degree of homogeneity in $\bk$, and thus does not depend on the modulus $k$. The integral over $k$ can therefore be done right away, resulting in the distribution
\begin{align}
\label{eq:intk}
\int_0^\infty \frac{\dd k\,k\, \ee^{\ii \bk\cdot\br'-a k|\bhk\cdot\bbm|}}{\bk\cdot\bbm+\ii 0^+}
&=\int_0^\infty \frac{\dd k\,\ee^{\ii k(\bhk\cdot\br'+\ii a|\bhk\cdot\bbm|)}}{\bhk\cdot\bbm+\ii 0^+}
=\frac{\ii}{(\bhk\cdot\bbm+\ii 0^+)(\bhk\cdot\br'+\ii a |\bhk\cdot\bbm|)}.
\end{align}
To deal with the remaining angular integral over $\bhk$ in (\ref{eq:beta0}), one introduces a rotated orthonormal basis $\{\bl,\bhk\}$, with components $\bl=(\cos\phi,\sin\phi)$ and $\bhk(\phi)=\partial\bl(\phi)/\partial\phi=(-\sin\phi,\cos\phi)$ in the fixed basis $\{\bbm,\bn\}$ (Fig.\ \ref{fig:fig1}b). By periodicity, the angular integral over the direction $\bhk$ in (\ref{eq:beta0}) is equivalent to one over $\phi$. This brings (\ref{eq:beta0}) down to the form
\begin{align}
\label{eq:beta0aaa}
\beta^0_{ij}(\br,t)
=-\frac{b_l}{2\pi}\avphi{
\frac{k_i G^0_{jp}(\bk,\bk\cdot\bv) (k_q c_{pqkl}n_k)-n_i\delta_{jl}}{(\bhk\cdot\bbm+\ii 0^+)(\bhk\cdot\br'+\ii a |\bhk\cdot\bbm|)}
}.
\end{align}
From definition (\ref{eq:modcmplx}), contractions involving the usual elastic tensor $c_{ijkl}$ can be expressed in terms of contractions with the modified, complex-valued one, $\widetilde{c}_{ijkl}$. Because $\bv\cdot\bn=0$, we have
\begin{align}
\label{eq:cctilde}
\hk_q c_{pqkl} n_k&=\lim_{\epsilon\to 0}(\hk n)_{\epsilon\,pl},
\end{align}
which is real-valued. Using this result and (\ref{eq:g0kv}) brings the numerator in (\ref{eq:beta0aaa}) in the form
\begin{align}
\label{eq:bracket0}
k_i G^0_{jp}(\bk,\bk\cdot\bv)(k_q c_{pqkl}n_k)-n_i\delta_{jl}
&=\lim_{\epsilon\to 0}\Re\left(\hk_i [(\hk\hk)_\epsilon^{-1}\cdot(\hk n)_\epsilon]_{jl}-n_i\delta_{jl}\right).
\end{align}
To carry out the average over $\phi$, the constant vector $\bn$ is decomposed over the rotated basis $\{\bl,\bhk\}$ as
\begin{align}
\label{eq:nlk}
\bn&=(\bn\cdot\bhk)\bhk+(\bn\cdot\bl)\bl.
\end{align}
After some simplifications involving the identities $\bv\cdot\bn=0$ and $\bhk\cdot\bl=0$, which hold by definition of the rotated basis, the following expansions obtain:
\begin{subequations}
\begin{align}
\label{eq:expannk}
(n\hk)_\epsilon&=(\bn\cdot\bhk)(\hk\hk)_\epsilon+(\bn\cdot\bl)(l\,\hk)_\epsilon,\\
\label{eq:expankn}
(\hk n)_\epsilon&=(\bn\cdot\bhk)(\hk\hk)_\epsilon+(\bn\cdot\bl)(\hk\,l)_\epsilon,\\
\label{eq:expannn}
(n n)_\epsilon&=(\bn\cdot\bhk)^2(\hk\hk)_\epsilon+(\bn\cdot\bhk)(\bn\cdot\bl)[(\hk\,l)_\epsilon+(l\,\hk)_\epsilon]
+(\bn\cdot\bl)^2(l\,l)_\epsilon.
\end{align}
\end{subequations}
Substituting expressions (\ref{eq:nlk}) and (\ref{eq:expankn}) into (\ref{eq:bracket0}) yields, after some cancellation of terms,
\begin{align}
\label{bracket0}
k_i G^0_{jp}(\bk,\omega)(k_q c_{pqkl}n_k)-n_i\delta_{jl}
&\mathop{=}_{\epsilon\to 0}(\bn\cdot\bl)\Re\left\{\hk_i[(\hk\hk)_\epsilon^{-1}\cdot(\hk\,l)_\epsilon]_{jl}
-l_i\delta_{jl}\right\}\nonumber\\
&=-(\bhk\cdot\bbm)\Re\left\{\hk_i[(\hk\hk)_\epsilon^{-1}\cdot(\hk\,l)_\epsilon]_{jl}-l_i\delta_{jl}\right\},
\end{align}
where the identity $\bn\cdot\bl=-\bhk\cdot\bbm$ ($=\sin\phi$) has been used. Inserting this expression into (\ref{eq:beta0aaa}) results in
\begin{align}
\label{eq:beta0angav}
\beta^0_{ij}(\br,t)
&=\frac{b_l}{2\pi}\avphi{\frac{\Re\left\{\hk_i[(\hk\hk)_\epsilon^{-1}\cdot(\hk\,l)_\epsilon]_{jl}-l_i\delta_{jl}\right\}}
{\bhk\cdot\br'+\ii a |\bhk\cdot\bbm|}},
\end{align}
where a factor $\bhk\cdot\bbm$ has been eliminated between the denominator (\ref{eq:intk}) and the numerator (\ref{bracket0}), the prescription $+\ii 0^+$ being irrelevant in this case. The real and imaginary parts of the denominator are separated as
\begin{align}
\frac{1}{\bhk\cdot\br'+\ii a |\bhk\cdot\bbm|}=\frac{\bhk\cdot\br'}{(\bhk\cdot\br')^2+a^2(\bhk\cdot\bbm)^2}-\frac{\ii a |\bhk\cdot\bbm|}{(\bhk\cdot\br')^2+a^2 (\bhk\cdot\bbm)^2}.
\end{align}
Thanks to this decomposition, we keep in the integrand only the even terms, which do not change their sign under the inversion symmetry $\phi\to\phi+\pi$ (\i.e., $\bhk\to -\bhk$ and $\bl\to -\bl$); the odd ones do not contribute. Thus, expression (\ref{eq:beta0angav}) takes the form
\begin{align}
\label{eq:beta0reim0}
\beta^0_{ij}(\br,t)
&=\frac{b_l}{2\pi}\Re\avphi{\frac{[(\hk\hk)_\epsilon^{-1}\cdot(\hk\,l)_\epsilon]_{jl}(\bhk\cdot\br')\hk_i}{(\bhk\cdot\br')^2+a^2 (\bhk\cdot\bbm)^2}}
-\frac{b_l}{2\pi}\avphi{\frac{(\bhk\cdot\br')l_i}{(\bhk\cdot\br')^2+a^2 (\bhk\cdot\bbm)^2}}\delta_{jl}.
\end{align}
Invoking next identity (\ref{eq:nnnmm1}) to simplify the leading angular integral yields
\begin{align}
\label{eq:beta0flat}
\beta^0_{ij}(\br,t)
&=-\frac{b_l}{2\pi}\Re\sum_{\alpha=1}^6 A^\alpha_{\epsilon\,j} L^\alpha_{\epsilon\,l}\avphi{\frac{p_\epsilon^\alpha(\phi)(\bhk\cdot\br')\hk_i }{(\bhk\cdot\br')^2+a^2(\bhk\cdot\bbm)^2}}
-\frac{b_l}{2\pi}\avphi{\frac{(\bhk\cdot\br')l_i}{(\bhk\cdot\br')^2+a^2(\bhk\cdot\bbm)^2}}\delta_{jl},
\end{align}
where subscripts have been introduced to emphasize that the quantities $\bA^\alpha$, $\bL^\alpha$, and $p^\alpha(\phi)$ now depend on $\epsilon$. Expression (\ref{eq:beta0flat}) reveals that the finite width $a$ has a twofold regularizing action on the angular averages: on the one hand, it sets the result to zero at the origin $r'=0$; on the other hand, it provides a regularization to handle the singularities at $\bhk\cdot\bhr'=0$. The Volterra limit $a\to 0^+$ is examined in Sec.\ \ref{sec:machcones} below.

To complete the calculation the following shorthand notations are used:
\begin{align}
\label{eq:shnot}
p^{0\alpha}_\epsilon&=p^\alpha_\epsilon(0),\qquad s_\alpha=\sign(\Im p^{0\alpha}_\epsilon),\qquad s_y=\sign y.
\end{align}
The main difference with the usual formalism (Sec.\ \ref{sec:constroh}) is that because the infinitesimal $\epsilon$ is always non-zero, we have $s_\alpha\not=0$ whatever the dislocation velocity, even though $p^{0\alpha}_\epsilon$ can become real-valued in the limit $\epsilon\to 0$ for supersonic velocities. This fact proves crucial in determining the Mach cone shapes in supersonic regimes, as will be seen in Sec.\ \ref{sec:machcones} below.

The angular integrals in (\ref{eq:beta0flat}) are carried out componentwise in the $\{\bbm,\bn\}$ basis, taking advantage of the explicit form (\ref{eq:ptan}) of the functions $p_\epsilon^\alpha$. To this aim we introduce the angle $\gamma$ such that
\begin{align}
\br'=r'[\cos\gamma\,\bbm+\sin\gamma\,\bn],\qquad \bhk\cdot\bhr'=\sin(\gamma-\phi).
\end{align}
In terms of the Cartesian coordinates $(x',y)=r(\cos\gamma,\sin\gamma)$ we obtain thus
\begin{subequations}
\label{eq:intflat}
\begin{align}
\avphi{\frac{p_\epsilon^\alpha(\phi)(\bhk\cdot\br')\bhk }{(\bhk\cdot\br')^2+a^2(\bhk\cdot\bbm)^2}}
&=\frac{1}{r}\avphi{\frac{p_\epsilon^\alpha(\phi)\sin(\gamma-\phi)}{\sin^2(\gamma-\phi)+(a/r)^2\sin^2\phi}
\left(
\begin{array}{c}
-\sin\phi\\
\cos\phi
\end{array}
\right)
}\nonumber\\
\label{eq:intflat1}
&=\frac{-s_y(|y|+a)\bbm+x'\bn}{x^{\prime 2}+(|y|+a)^2}-\frac{1}{\ii}\frac{\bbm+p_\epsilon^{0\alpha}\bn}{s_\alpha(x'+p_\epsilon^{0\alpha}y)+\ii s_y a},\\
\label{eq:intflat2}
\avphi{\frac{(\bhk\cdot\br')\bl}{(\bhk\cdot\br')^2+a^2(\bhk\cdot\bbm)^2}}
&=\frac{1}{r}\avphi{\frac{\sin(\gamma-\phi)}{\sin^2(\gamma-\phi)+(a/r)^2\sin^2\phi}
\left(
\begin{array}{c}
\cos\phi\\
\sin\phi
\end{array}
\right)
}
=\frac{s_y(|y|+a)\bbm-x' \bn}{x^{\prime 2}+(|y|+a)^2}.
\end{align}
\end{subequations}
Inserting these results into (\ref{eq:beta0flat}) and invoking the completeness relation (\ref{eq:cAA}${}_2$), we conclude that
\begin{align}
\label{eq:beta0flat1}
\beta^{0}_{ij}(\br,t)
&=\Re\frac{1}{2\ii\pi}\sum_{\alpha=1}^6 \frac{m_i+p_\epsilon^{0\alpha} n_i}{s_\alpha(x'+p_\epsilon^{0\alpha} y)+\ii s_y a}A^\alpha_{\epsilon\,j} L^\alpha_{\epsilon\,l} b_l.
\end{align}
For definiteness, the calculation of the first component (i.e., along $\bbm$) of integral (\ref{eq:intflat1}) ---by contour integration on the unit circle, as is usual within the Barnett-Lothe approach--- is explained in detail in \ref{sec:appflat}. The other integrals in this section are computed likewise.

We turn next to the ``propagator'' part, which by (\ref{eq:blineD}) and (\ref{eq:betapflat}), and after use of (\ref{eq:intk}), reads
\begin{align}
\beta^D_{ij}(\br,t)&=\int\frac{\dd^2\!k}{(2\pi)^2} B^D_{ijkl}(\bk,\bk\cdot\bv)\beta^{\rm p}_{kl}(\bk)
\ee^{\ii \bk\cdot\br'}
=\frac{\ii\,b_l}{2}\int\frac{\dd^2\!k}{(2\pi)^2}\frac {k_i D_{jp}(\bk,\bk\cdot\bv)(k_q c_{pqkl}n_k)}{\bk\cdot\bbm+\ii 0^+}
\ee^{\ii \bk\cdot\br'-ak|\bhk\cdot\bbm|}\nonumber\\
\label{eq:betaD}
&=-\frac{b_l}{2(2\pi)}\avphi{\frac {k_i D_{jp}(\bk,\bk\cdot\bv)(k_q c_{pqkl}n_k)}{(\bhk\cdot\bbm+\ii 0^+)(\bhk\cdot\br'+\ii a |\bhk\cdot\bbm|)}}.
\end{align}
From expression (\ref{eq:dkv}) of $\sfD(\bk,\bk\cdot\bv)$, we have
\begin{align}
\frac{1}{2}k_i D_{jp}(\bk,\bk\cdot\bv) (k_q c_{pqkl}n_k)
&=\ii \hk_i \Im [(\hk\hk)_\epsilon^{-1}]_{jp}(k_q c_{pqkl}n_k)\sign(\bhk\cdot\bbm).
\end{align}
so that, by substitution into (\ref{eq:betaD}),
\begin{align}
\beta^D_{ij}(\br,t)
&=-\ii\frac{b_l}{2\pi}\avphi{\frac {\hk_i \Im [(\hk\hk)_\epsilon^{-1}]_{jp}(k_q c_{pqkl}n_k)\sign(\bhk\cdot\bbm)}{(\bhk\cdot\bbm+\ii 0^+)(\bhk\cdot\br'+\ii a |\bhk\cdot\bbm|)}}.
\end{align}
Employing the same means as above, this expression reduces to
\begin{align}
\label{eq:betaDflat}
\beta^{D}_{ij}(\br,t)
&=-a\frac{b_l}{2\pi}\Im\sum_{\alpha=1}^6 A_{\epsilon\,j}^\alpha L_{\epsilon\,l}^\alpha\avphi{\frac{p_\epsilon^\alpha(\phi)(\bbm\cdot\bhk)\hk_i }{(\bhk\cdot\br')^2+a^2(\bhk\cdot\bbm)^2}}.
\end{align}
The angular integral evaluates to
\begin{align}
\label{eq:angintDflat}
\avphi{\frac{p^\alpha(\phi)(\bbm\cdot\bhk)\bhk}{(\bhk\cdot\br')^2+a^2|\bhk\cdot\bbm|^2}}
&=-\frac{1}{r^2}\avphi{\frac{p_\epsilon^\alpha(\phi)\sin\phi}{\sin^2(\gamma-\phi)+(a/r)^2\sin^2\phi}
\left(
\begin{array}{c}
-\sin\phi\\
\cos\phi
\end{array}
\right)
}\nonumber\\
&=-\frac{1}{a}\left[\frac{-s_y x'\,\bbm+(|y|+a)\bn}{x'^2+(|y|+a)^2}+\frac{\bbm+p_\epsilon^{0\alpha}\bn}{s_y(x'+p^0_\alpha y)+\ii s_\alpha a}
\right].
\end{align}
Upon substituting into (\ref{eq:betaDflat}), we obtain thus
\begin{align}
\beta^{D}_{ij}(\br,t)
&=\Im\frac{1}{2\pi}\sum_{\alpha=1}^6\left[\frac{-s_y x'\,m_i+(|y|+a)n_i}{{x'}^2+(|y|+a)^2}+\frac{m_i+p_\epsilon^{0\alpha} n_i}{s_y(x'+p_\epsilon^{0\alpha} y)+\ii s_\alpha a}\right]A_{\epsilon\,j}^\alpha L_{\epsilon\,l}^\alpha b_l.
\end{align}
However, using once more the completeness relation (\ref{eq:cAA}.2), we observe that the first term within brackets is real-valued and does not contribute to the result due to the leading $\Im$ operator. Therefore,
\begin{align}
\label{eq:betaDflat1}
\beta^{D}_{ij}(\br,t)
&=\Im\frac{1}{2\pi}\sum_{\alpha=1}^6\frac{m_i+p_\epsilon^{0\alpha} n_i}{s_y(x'+p_\epsilon^{0\alpha} y)+\ii s_\alpha a} A_{\epsilon\,j}^\alpha L_{\epsilon\,l}^\alpha b_l.
\end{align}

Eventually, adding Eqs.\ (\ref{eq:beta0flat1}) and (\ref{eq:betaDflat1}) and carrying out a few obvious reorganizations yields the total elastic distortion in the form
\begin{align}
\label{eq:betaflat}
\beta_{ij}(\br,t)&=\beta^{0}_{ij}(\br,t)+\beta^{D}_{ij}(\br;t)
=\frac{1}{2\pi}\Im\sum_{\alpha=1}^6
\frac{(s_\alpha+s_y)(m_i+p_\epsilon^{0\alpha} n_i)}{(x'+p_\epsilon^{0\alpha} y)+\ii s_y s_\alpha a}A_{\epsilon\,j}^\alpha L_{\epsilon\,l}^\alpha b_l.
\end{align}

\section{The Volterra limit and Mach cones}
\label{sec:machcones}
\subsection{Mach cones as Dirac measures}
\label{sec:diraclines}
The distortion components (\ref{eq:betaflat}) in the Volterra limit $a\to 0$ are easily deduced. To this purpose, it is necessary to distinguish between real and complex roots $p_\epsilon^{0\alpha}$ in the limit $\epsilon\to 0$. When $p_\epsilon^{0\alpha}$ is such that its limit $p^{0\alpha}\equiv\lim_{\epsilon\to 0}p_\epsilon^{0\alpha}$ is real, we get from the Sokhotski-Plemelj formula (\ref{eq:sk})
\begin{align}
\label{eq:skp}
\lim_{\epsilon\to 0}\frac{1}{x'+p_\epsilon^{0\alpha}y+\ii s_y s_\alpha 0^+}&=\pv\frac{1}{x'+p^{0\alpha}y}-\ii\pi s_y s_\alpha\delta(x'+p^{0\alpha}y),
\end{align}
where, because it is defined as a limit, $s_\alpha=\lim_{\epsilon\to 0}\sign\Im p_\epsilon^{0\alpha}$ must be considered as nonzero even though the limit $p^{0\alpha}$ is real [see remark after Eq.\ (\ref{eq:shnot})]. Substituting (\ref{eq:skp})  into (\ref{eq:betaflat}), and observing that $s_y s_\alpha(s_y+s_\alpha)=s_y+s_\alpha$ yields the distributional expression (where the superscript V on $\beta_{ij}$ stands for `Volterra')
\begin{align}
\label{eq:betavolt}
\beta^{{\rm V}}_{ij}(\br,t)&=
\frac{1}{2\pi}\Im\sum_{\alpha}\pv\frac{(s_\alpha+s_y)(m_i+p^{0\alpha} n_i)}{x'+p^{0\alpha} y}A_{\epsilon\,j}^\alpha L_{\epsilon\,l}^\alpha b_l\nonumber\\
&{}-\frac{1}{2}\sum_{\alpha}{}^{'}(s_\alpha+s_y)(m_i+p^{0\alpha} n_i)\Re\left[A_{\epsilon\,j}^\alpha L_{\epsilon\,l}^\alpha\right] b_l \delta(x'+p^{0\alpha} y).
\end{align}
The principal-value prescription within the first sum is necessary only for the subset of $\alpha$ values such that $\Im p^{0\alpha}=0$, and can be omitted whenever $p^{0\alpha}$ has a finite imaginary part since the latter prevents the denominator from vanishing. The `prime' restricts the second sum to the same subset of $\alpha$ values.

Mach cones are characterized as follows. The rightmost sum in (\ref{eq:betavolt}) features Dirac measures supported by straight lines of equations $y=-x'/p^{0\alpha}$, which represent the Mach cones. Each V-shaped cone is made of two half-lines (`branches'), and thus is associated with two different eigenvalues $p^\alpha$. Selection of one particular cone branch $\alpha$ is done via its factor $(s_\alpha+s_y)$. For instance, the branch of a Mach cone in the upper half-plane $y>0$ is such that $s_\alpha=+1$, because $s_\alpha+s_y=2\not=0$ if $y>0$ and zero otherwise; the same holds with all signs reversed for the lower branch in the half-plane $y<0$. Moreover, the polarization scalar product $\bL_{\epsilon}^\alpha\cdot\bb$ makes the intensity and sign of the elastic fields of each branch depend on the dislocation character. In particular, the ``ghost branches'' for which this scalar product vanishes are extinguished and do not contribute to the field even though they exist ---so to say--- geometrically speaking.

\subsection{The classical subsonic result}
In the subsonic range $|v|<c_{\rm l}$ where all the $p^{0\alpha}$ have non-zero imaginary part, Eq.\ (\ref{eq:betavolt}) reduces to
the `classical' result \cite{LOTH92b,HIRT82} for a Volterra dislocation. Indeed, we have then
\begin{align}
\label{eq:betavoltsub0}
\beta^{{\rm V}}_{ij}(\br,t)&=\frac{1}{2\pi}\Im\sum_{\alpha=1}^6\frac{(s_\alpha+s_y)(m_i+p^{0\alpha} n_i)}{x'+p^{0\alpha} y}A_{\epsilon\,j}^\alpha L_{\epsilon\,l}^\alpha b_l.
\end{align}
Remembering that $s_\alpha=\sign\Im p^{0\alpha}$, our numbering conventions and the complex-conjugacy properties of the eigenvalues and eigenvectors (see Section \ref{sec:constroh}) imply that
\begin{align}
\beta^{{\rm V}}_{ij}(\br,t)
&=\frac{1}{2\pi}\Im\left[\sum_{\alpha=1}^3\frac{(1+s_y)(m_i+p^{0\alpha} n_i)}{x'+p^{0\alpha} y}A_{\epsilon\,j}^\alpha L_{\epsilon\,l}^\alpha+\overline{\sum_{\alpha=1}^3\frac{(-1+s_y)(m_i+p^{0\alpha} n_i)}{x'+p^{0\alpha} y}A_{\epsilon\,j}^\alpha L_{\epsilon\,l}^\alpha}\right] b_l\nonumber\\
&=\frac{1}{\pi}\Im\sum_{\alpha=1}^3\frac{(m_i+p^{0\alpha} n_i)}{x'+p^{0\alpha} y}A_{\epsilon\,j}^\alpha L_{\epsilon\,l}^\alpha b_l
=\frac{1}{2\ii\pi}\sum_{\alpha=1}^6 s_\alpha\frac{(m_i+p^{0\alpha} n_i)}{x'+p^{0\alpha} y}A_{\epsilon\,j}^\alpha L_{\epsilon\,l}^\alpha b_l,
\end{align}
which is the classical result.

\subsection{Geometric constructions: forward and backward Mach cones}
Mal\'en \cite{MALE70d} provided an explanation of Mach cones from a geometric construction based on the section of the \emph{slowness surface} in the sagittal plane. As we shall shortly see, the argument is incomplete as it only involves phase velocities and does not allow one to explain Payton's `backward' cones \cite{PAYT95}(see Introduction). The geometrical construction that provides the explanation involves group velocities.

First, we briefly review the definition of slowness and group-velocity surfaces (see, e.g., \cite{AULD73,ROY00a} for details), also known as \emph{ray surfaces} in the literature (e.g., \cite{SPIE10}). With definition (\ref{eq:omegadef}) of $\Omega$, the dispersion equation $\smash{\Omega(\bk,\omega)=k^6\Omega(\bhk,\omega/k)=0}$ is solved for the (positive) phase velocities $\smash{c(\bhk)\defi\omega/k}$ as a function of the wave direction $\smash{\bhk}$ in the sagittal plane. The slowness vectors are $\bs_\alpha(\bhk)\defi\smash{c_\alpha^{-1}(\bhk)\bhk}$. Parametric plots of the endpoints of the vectors $\bs_\alpha(\bhk)$, using $\bhk$ as a parameter, define slowness surfaces, in which the radius vector lies along $\bhk$.
By contrast, group velocities are defined as
\begin{align}
\label{eq:gvdef}
\bv^{\rm g}_\alpha(\bhk)\defi\bnabla_{\bk}\omega(\bk)=-\left.\frac{\bnabla_{\bk}\Omega(\bk,\omega)}{\partial_\omega\Omega(\bk,\omega)}\right|_{\omega=c_\alpha(\bhk)k}.
\end{align}
Parametric plots of the endpoints of the vectors $\bv^g_\alpha$ using $\bhk$ as a parameter define group-velocity surfaces. The latter represent the positions at time $t=1$ s of wavefronts emitted from the origin at $t=0$. Group-velocity surfaces are sometimes also called 'energy-velocity surfaces'. The velocity of energy transport in a plane wave is indeed equal to its group velocity. Group and phase velocities are different from one another in anisotropic media due to directional interference effects, which makes anisotropic media dispersive \cite{LIGH60,STRO62}. The vector $\bv^{\rm g}_\alpha(\bhk)$ lies in the direction normal to the slowness surface at $\bhk$ and, save for special directions of symmetry, is not directed along $\bhk$. Quite generally,
\begin{align}
\norm{\smash{\bv^{\rm g}_\alpha(\bhk)}}\geq\smash{\bv^{\rm g}_\alpha(\bhk)}\cdot\bhk=c_\alpha(\bhk).
\end{align}

\begin{figure}[!ht]
\centering
\includegraphics[width=12cm]{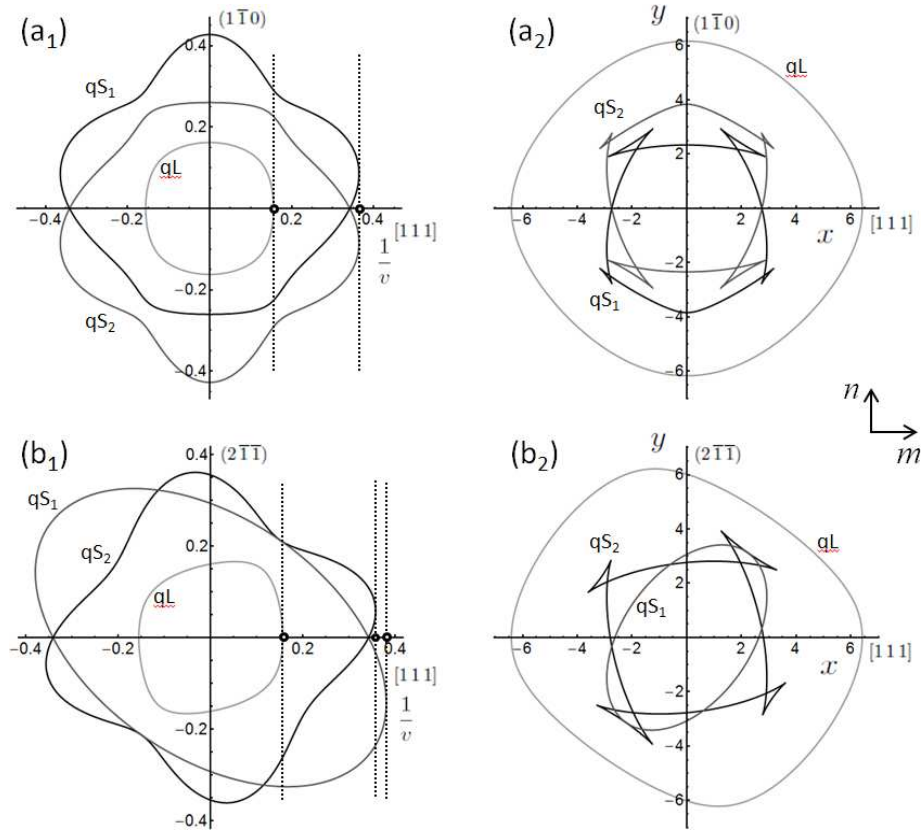}
\caption{\label{fig:fig2} Slowness and group-velocity surfaces for Fe. (a${}_{1}$) and (b${}_{1}$): sections of slowness surfaces in the sagittal plane $(\bbm,\bn)$ (in units of $10^{-3}$ s\,m${}^{-1}$). Open dots: limiting values for $1/v$, where $v$ is the source velocity; (a${}_{2}$) and (b${}_{2}$): sections of group-velocity surfaces in the sagittal plane (in units of $10^3$  m\,s${}^{-1}$). (a${}_{1,2}$): slip system $\bbm=[1\,1\,1]$, $\bn=(1\,\overline{1}\,0)$ (dislocation direction $\bxi=[1\,1\,\overline{2}]$); (b${}_{1,2}$): slip system $\bbm=[1\,1\,1]$, $\bn=(2\,\overline{1}\,\overline{1})$ (dislocation direction $\bxi=[0\,1\,\overline{1}]$).}
\end{figure}

Figures \ref{fig:fig2}(a${}_1$) and (b${}_1$) represent sections of slowness surfaces in Fe,\footnote{Elastic constants (GPa) $C_{11}=226.0$, $C_{12}=140.0$, $C_{44}=166.0$, mass density $\rho=7.8672$ g/cm${}^3$, and lattice parameter $a_0=0.287$ nm at 298 K \cite{LIDE10}.} in the sagittal plane of the two different slip systems (see legend). In spite of intersection points, the three individual branches of solution are unambiguously identified using the fact that they are smooth functions of $\bhk$. Surfaces ({\rm qS}${}_{1,2}$) in the figure are of quasi-shear character (shear polarization with admixture of longitudinal polarization), and the inner one (qL) is quasi-longitudinal (longitudinal polarization with admixture of shear polarization). Intersections of slowness curves with vertical lines of abscissa $1/v$ have coordinates $\bs_\alpha=(1/v,p^{0\alpha}/v)$, which provides the connection with the Stroh formalism \cite{MALE70d}. Such intersections exist only if their associated $p^{0\alpha}$ is real. This is possible only for $v$ above certain bulk limiting velocities of the problem (open dots in the figure) for which the vertical lines are tangent to the slowness curves (represented as dashed in the figures). Thus, limiting velocities are $v_{\rm L}\simeq 2.75\,10^3$, and $6.41\,10^3$ m\,s${}^{-1}$ for the slip system in Fig.\ \ref{fig:fig2}(a${}_1$), and $v_{\rm L}\simeq 2.63\,10^3$, $2.80\,10^3$, and $6.41\,\,10^3$ m\,s${}^{-1}$ for the slip system in Fig.\ \ref{fig:fig2}(b${}_1$). Whereas the generic case, illustrated in Fig.\ \ref{fig:fig2}(b${}_1$) is that of three limiting velocities, the case of Fig.\ \ref{fig:fig2}(a${}_1$) is degenerate since it involves only two velocities. A similar degeneracy arises in isotropic media where the three slowness curves are perfect circles, with the two shear curves superimposed.

Figures \ref{fig:fig2}(a${}_2$) and (b${}_2$) represent sections of group-velocity surfaces in Fe in the sagittal plane of two slip systems considered. Cusps in group-velocity surfaces arise from non-convex parts in slowness surfaces (here, in ${\rm qS}$ branches). When interpreted in units of meters, group-velocity surfaces provide the shape of wavefronts at $t=1$~s emitted at $t=0$ by a source at the origin of coordinates \cite{AULD73,ROY00a}.

Now, let $\mathcal{R}_0$ be this set of group-velocity surfaces centered on the origin. The well-known Huygens construction \cite{FREU72,KAOU08,MARK08} for dislocations in isotropic media (where the situation is simpler because phase and group velocities coincide) can be generalized to anisotropic media, using group-velocity surfaces to build Mach cones as envelopes of the collection of wavefronts emitted over time \cite{SPIE10}. Thus, let $T_{\bs}$ be the translation operator that translates a set of vectors by a vector $\bs$. The collection of wave fronts emitted by a moving dislocation located at $\br=\mathbf{0}$ at time $\tau=t$ can then be written as
\begin{align}
\mathcal{W}(t)=\cup_{0\leq \tau\leq t}(t-\tau) T_{\bs_t(\tau)}\mathcal{R}_0,
\end{align}
which is the union over time of sets of ray surfaces. Each set is built from $\mathcal{R}_0$, translated by the retarded position vector $\bs_t(\tau)$ of the dislocation (to account for the motion of the emission point of the pulses), and expanded homothetically by a factor $(t-\tau)$ (to account for pulse expansion by wave propagation). Mach cones are envelopes of this set (i.e., caustics, where radiated energy is concentrated) \cite{KAOU08}. For the uniform motion at velocity $\bv$ considered here one has $\bs_t(\tau)=\bv(t-\tau)$. However, the construction, that rests on the concept of retarded fields, holds for non-uniform motion as well \cite{FREU72}. As it is only of geometric nature, it provides no information about the intensity of cone branches, some of which may vanish for reasons of polarization as noted in Sec.\ \ref{sec:diraclines}.

\begin{figure}[!ht]
\centering
\includegraphics[width=12cm]{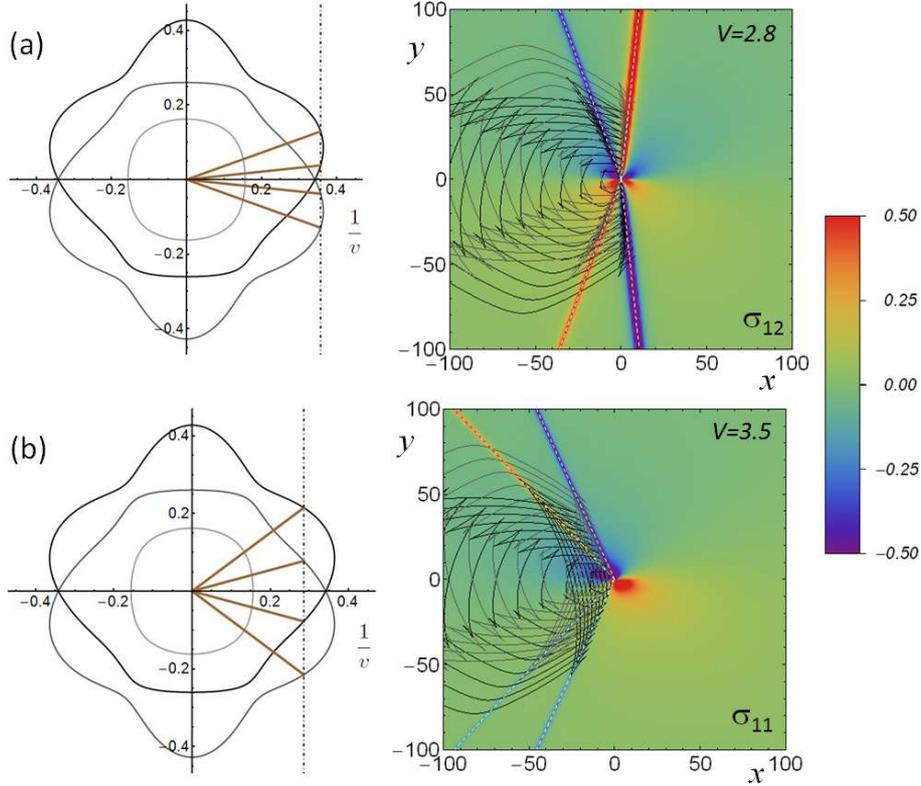}
\caption{\label{fig:fig3} Edge dislocation in Fe, with $\bb=(a_0/2)\moy{1 1 1}$, $\bn=\{1,\overline{1}\ 0\}$ for the two velocities indicated (units of $10^3$ m\,s${}^{-1}$). Left: solutions on the slowness surfaces; right: selected full-field stress components computed from Eq.\ (\ref{eq:betaflat}), with the Huygens construction superimposed. The half core width is $a=1.355\,d$ in (a) and $a=0.967\,d$ in (b) \cite{PELL17}. Distances are measured in units of the interplane distance $d$ for this glide system. For better display fields have been cut-off as indicated in the bar legend. White dashed lines drawn upon the cones are the theoretical loci of cone-branch lines (see Sec.\ \ref{sec:diraclines}). For better display, the construction only involves a restricted number of wavefronts built from the quasi-shear branches ${\rm qS}_{1,2}$ of Fig.\ \ref{fig:fig2}(a${}_2$) (the only relevant branches for Mach cones, at the velocities considered). (For a color version of the figure, the reader is referred to the web version of this article.)}
\end{figure}

The construction is illustrated in Fig.\ \ref{fig:fig3} where stress components computed from the field formula (\ref{eq:betaflat}) are displayed for two velocities in a full-field representation (right), together with the corresponding solutions on the slowness surfaces (left). The velocity in (a), $v=2.8\,10^3$ m\,s${}^{-1}$ is such that one forward and one backward cones are present. In (b), the velocity is higher, and two forward cones are generated. On the full-field plots have been superimposed the Huygens construction from the group-velocity surfaces of Fig.\ \ref{fig:fig2}, as well as cone lines of equations $x'+p^{0\alpha} y=0$ deduced form the theoretical expression (\ref{eq:betavolt}). The former perfectly reproduces the latter in agreement with the full-field plots. This triple comparison makes clear that the envelope of the cusps of the group-velocity surfaces of Fig.\ \ref{fig:fig2} does \emph{not} determine the Mach cone opening angle, as those endpoints do not pertain to any caustic in general: from a mathematical standpoint the cone opening angle is uniquely defined from the above linear equation, which clarifies some ambiguities in recent interpretations of the Huygens construction in anisotropic media \cite{SPIE10}.

\begin{figure}[!ht]
\centering
\includegraphics[width=10cm]{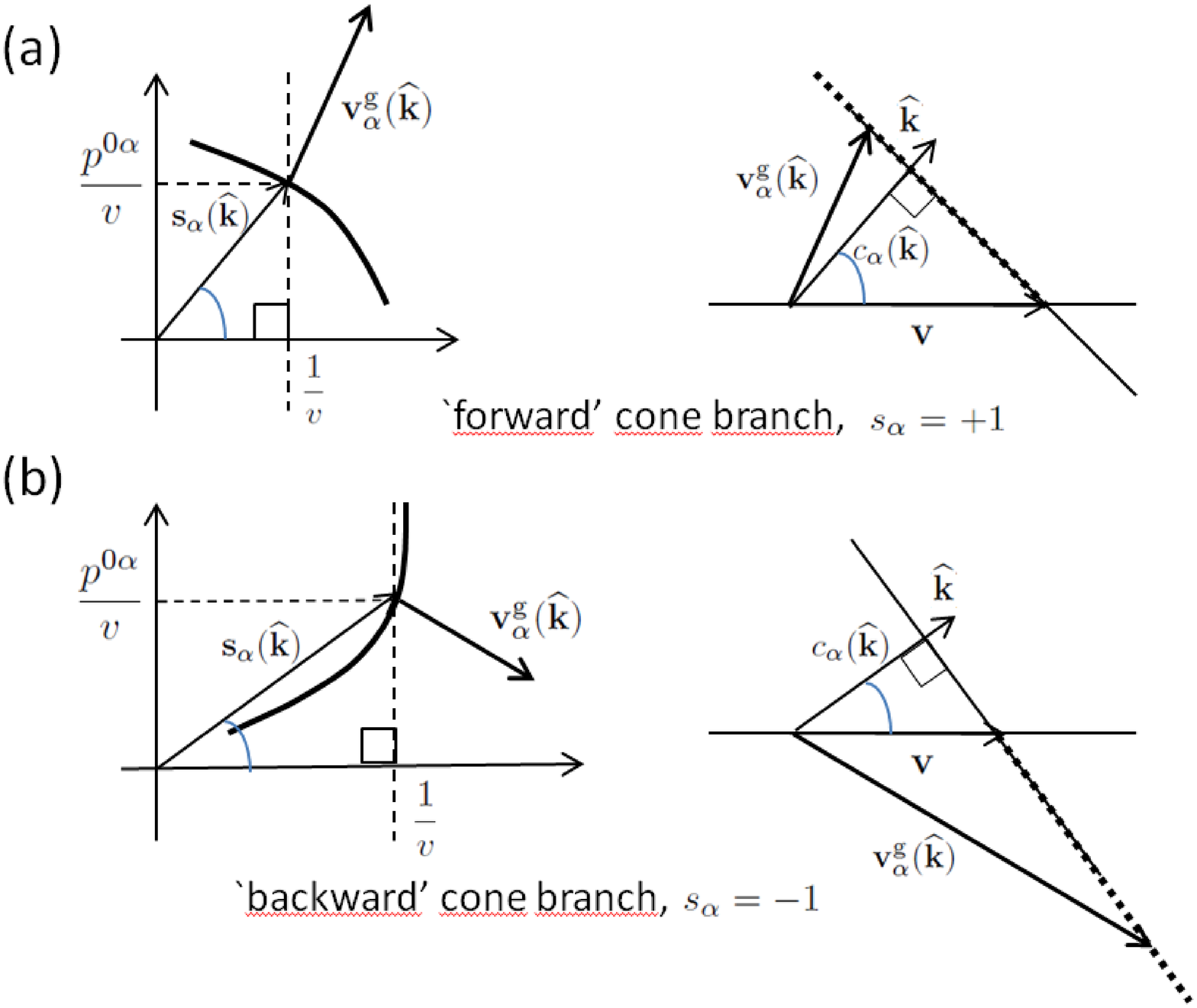}
\caption{\label{fig:fig4} Geometric constructions for forward and backward Mach cones including group-velocity considerations. (a) `normal' (forward-cone) case; (b) `anomalous' (backward-cone) case. Left: constructions from a piece of the section of the slowness surface. Right: corresponding constructions in physical space, with backward (top) and forward (bottom). The resulting half-branches of the Mach cones are represented as dashed lines.}
\end{figure}

Fig.\ \ref{fig:fig3}(a) correlates the existence of backward cones with the fact that one pair of solutions deduced from the slowness surfaces involve normals (group-velocity vectors) pointing downwards in the upper half plane, and upwards in the lower half-plane. Fig.\ \ref{fig:fig4}, which extends Mal\'en's construction to account for group-velocity effects \cite{KOIZ02} illustrates this in detail. The angle indicated is the same in the slowness-surface construction (left) and the vector construction in the physical space (right). The half-cone branches are indicated as dashed lines. The bottom construction shows that while the wave vector $\bhk$ points upwards, the group velocity points downwards, and is responsible for the physical backward Mach cone branch in the lower half-plane. In both cases, the following relation is obeyed, with $\bv=v\bbm$:
\begin{align}
c_\alpha(\bhk)&=\bv\cdot\bhk=\bv^g_\alpha(\bhk)\cdot\bhk.
\end{align}

Examination of  Eq.\ (\ref{eq:skp}) reveals that for this effect to be reproduced by the Mach-cone (Dirac) part in this equation, it is necessary that $s_\alpha=\lim_{\epsilon\to 0}\sign\Im p^{0\alpha}_\epsilon$ $=\pm 1$ be equal to the sign of $\bv^g\cdot\bn$. This property we demonstrate as follows. First, by setting $\bk=\bbm+p\bn$, and $\omega=v$,\footnote{Both these quantities should be multiplied by some arbitrary dimensioning factor, identical for both, here assumed equal to 1 m${}^{-1}$.} it is easily verified that
\begin{align}
\Delta(p,v)&=\Omega(\bk,\omega),
\end{align}
which makes explicit the connection between the dispersion equation, see Eq.\ (\ref{eq:omegadef}), and the Stroh eigenvalue equation (\ref{eq:pv}). Note in passing that solutions are such that $c=\omega/k=v/\sqrt{1+p^2}$, in agreement with the geometric constructions of Fig.\ \ref{fig:fig4} (left) where $\bs=\bhk/c$. Now, adding a small imaginary part $\delta v=\ii\epsilon$ to $v$ ($\epsilon>0$), any real solution $p(v)$ of the equation $D(p,v)=0$ varies by an amount $\delta p$, and $\omega$ and $\bk$ vary by respective amounts $\delta\omega=\delta v$ and $\delta\bk=\delta p\,\bn$. Thus,
\begin{align}
0=\Delta(p+\delta p,v+\delta v)
&=\Omega(\bk+\delta\bk,\omega+\delta\omega)\nonumber\\
&\simeq\Omega(\bk,\omega)+\bnabla_{\bk}\Omega(\bk,\omega)\cdot\bn\delta p+\partial_\omega\Omega(\bk,\omega)\delta v.
\end{align}
Then, because $p$ and $v$ are such that $\Delta(p,v)=\Omega(\bk,\omega)=0$, one deduces that
\begin{align}
\delta p
&=-\frac{\partial_\omega\Omega(\bk,\omega)}{\bnabla_{\bk}\Omega(\bk,\omega)\cdot\bn}\delta v
=\frac{\ii\epsilon}{\bv^g_{\bk}\cdot\bn},
\end{align}
where definition (\ref{eq:gvdef}) of the group velocity has been used. So, the infinitesimal imaginary part of $p$ generated by the adjunction of $\ii\epsilon$ to the velocity in the Stroh formalism, to account for causality, has indeed the necessary sign.

From sections of slowness surfaces, it easy to assess the range of velocities for which backward cones can be present: it is simply the one where there exists normals to the slowness plots, that point towards the horizontal axis. For instance, in the two cases of Fig.\ \ref{fig:fig2}, those ranges are comprised between the lower bulk limiting velocity, and the velocity where the two quasi-shear slowness branches {\rm qS}${}_{1,2}$ intersect mutually on the horizontal axis; namely, $2.75 \leq v\leq 2.93$ in case (a${}_1$), and $2.63\leq v\leq 2.93$ in case (b${}_1$)  (in units of $10^3$ m\,s${}^{-1}$).

Leaving it to the reader, the same analysis as in Sec.\ \ref{sec:diraclines} could be carried out separately on the reactive and radiative parts of the distortion field, Eqs.\ (\ref{eq:beta0flat1}) and (\ref{eq:betaDflat1}), respectively. One would see that, individually, each of those parts produces for faster-than-wave motion X-shaped wavefronts involving a-causal cone branches, instead of V-shaped ones. However, the unphysical branches vanish by sign compensation in the sum of both terms, Eq.\ (\ref{eq:betavolt}).

\section{Concluding discussion}
\label{sec:concl}
To conclude, we summarize our results and put them into perspective. First, we showed that the Stroh formalism can be very easily extended to supersonic velocities by means of an analytic continuation to complex values of the algebraic velocity $v$, with infinitesimal positive imaginary component. This device was discovered in former investigations by Pellegrini of the isotropic theory of the equation of motion of dislocations \cite{PELL12,PELL14}. However that first proof stemmed from a somewhat \textit{ad hoc} argument, as it resulted from comparing formulas \cite{ROSA01} previously obtained by considering separately the subsonic and supersonic cases. Instead, by tracing back the analytic continuation to the radiation condition that defines the anisotropic Green functions, the present work gives the method a firm physical basis, and puts it into a far broader perspective, paving the way, e.g., for studying supersonic regimes in the anisotropic Weertman equation. Concretely, it is no more necessary to derive field-related velocity-dependent theoretical expressions separately in the subsonic and supersonic regimes to cover the whole velocity range, as was previously done. It now merely suffices to compute subsonic expressions, and to continue them by means of the replacement $v\to v+\ii\epsilon$, which is most easily done in numerical computations by making $\epsilon$ a very small number. In this way, an expression valid for \emph{all} velocities is obtained. Adams \cite{ADAM01} came close to the point in 2001, in the context of a Weertman-type equation for slip-pulse propagation at the interface between two different isotropic media. He noticed that unique expressions of the coefficients of the governing equation apply indifferently to subsonic or intersonic velocities, provided that all `relativistic' terms involved, of the type $\sqrt{1-v^2/c^2}$ (where $c$ is any of the four wavespeeds involved) for $|v|<c$, are replaced for $|v|>c$ by $-\ii\sign(v)\sqrt{v^2/c^2-1}$, which he rightfully identified as a `radiation condition' without however giving any further explanation. In fact, as remarked by Pellegrini \cite{PELL12} (see Sec.\ 7 and Appendix A in that reference), the following complex identity valid for the principal determination of the square root, namely, $\sqrt{-z^2}=-\ii\sign(\Im z) z$ for $z\in\mathbb{C}\backslash\mathbb{R}$ implies that, for $\epsilon>0$,
\begin{align}
\sqrt{1-(v+\ii\epsilon)^2/c^2}=\sqrt{-[\sqrt{(v+\ii\epsilon)^2/c^2-1}]^2}=-\ii\sign(v)\sqrt{(v+\ii\epsilon)^2/c^2-1}.
\end{align}
We immediately deduce that, with $\theta(x)$ the Heaviside unit-step function,
\begin{align}
\lim_{\epsilon\to 0}\sqrt{1-(v+\ii\epsilon)^2/c^2}=\sqrt{1-v^2/c^2}\,\theta(1-|v|/c)-\ii\sign(v)\sqrt{v^2/c^2-1}\,\theta(|v|/c-1).
\end{align}
Thus, in the isotropic case, carrying out the analytic continuation is equivalent to implementing Adams's radiation-condition prescription. In the anisotropic case where, employing Stroh's formalism, the functions involved must ultimately be computed numerically (except in high-symmetry cases), the analytic continuation generalizes Adams's prescription in an `automatic' way, alleviating the need to \emph{actually know} these functions in closed analytical form.

Second, we derived an explicit expression for the distortion field of an Eshelby dislocation in an anisotropic medium, Eq.\ (\ref{eq:betaflat}), from which the stress field immediately follows. The expression reduces to the classical result in the Volterra limit and in the subsonic regime. However, it was extended to supersonic velocities by virtue of the above analytic continuation. It should be noted, however, that the occurrence of the sign $s_y=\sign(y)$ in Eq.\ (\ref{eq:betavolt}) is non-trivial. In the Volterra limit, this allowed us to derive analytically the Mach cone structure. Keeping instead the core width finite, we could effectively evaluate numerically the Mach cones from Eq.\ (\ref{eq:betaflat}), which was illustrated by full-field plots. In this respect, the present work can be seen as a continuation of our previous efforts relative to isotropic media \cite{PELL15,LAZA16}.

Third, Payton's `backward' Mach cones were given an explanation, and a simple criterion was given to determine from slowness surfaces the velocity range in which they show up. The range is simply the one for which waves have a normal to the slowness surface (proportional to the group velocity) that points towards the abscissa axis. There exists a well-known close connection \cite{STRO62,BARN73a,BARN73b,ALSH81} between the problem of a uniformly-moving dislocation in an anisotropic media, and the theory of surface waves and of wave reflection at interfaces in anisotropic media (see, e.g., \cite{BARN00,LOTH09,FAVR11} for recent reviews). In the latter context, it is remarked that waves with such normals have been interpreted as incident waves onto the interface \cite{ALSH81}. Our analysis of `backward' cones shows that such waves can also be radiated away \emph{from} the glide plane (but the side opposite as the usual one), thus offering a new perspective on their practical significance. The very natural question as to whether fully-developed `backward' March cones could effectively be observed in atomistic simulations or in finite-element calculations is an issue beyond the scope of this paper. To answer it would require at least investigating the stability of such steady motions \cite{MALE70d,ROSA01}. This problem is connected with the determination of the width of the Eshelby dislocation as a function of the velocity \cite{ESHE49a,WEER69b,ROSA01,PELL14}, and will be examined elsewhere \cite{PELL17}.

\section*{Acknowledgments}
The author thanks A.\ Vattr\'e for discussions about the Stroh formalism in the preliminary steps.

\appendix
\section{Useful distributions}
\label{sec:distr}
Some useful distributions employed in the paper are briefly recalled. First, the well-known Sokhotski-Plemelj formula is
\begin{align}
\label{eq:sk}
\frac{1}{x\pm\ii 0^+}=\pv\frac{1}{x}\mp\ii\pi\delta(x).
\end{align}
Taking its derivative with respect to $x$ gives
\begin{align}
\label{eq:dsk}
\frac{1}{(x\pm\ii 0^+)^2}=\Pf\frac{1}{x^2}\pm\ii\pi\delta'(x).
\end{align}
which can be generalized to any number of successive differentiations (e.g., \cite{GELF64}, p.\ 94).

The interesting identity (\ref{eq:curious}) is proven as follows. Introduce the integral
\begin{align}
I_\epsilon(\br)=\Re\int_0^{2\pi}\frac{\dd\phi}{(\bhk\cdot\br+\ii\epsilon)^2},
\end{align}
where $\phi$ is the angle between $\bhk$ and $\br$.  By (\ref{eq:dsk}), the limit of $I_\epsilon(\br)$ as $\epsilon\to 0$ is equal to the integral in the left-hand side of Eq.\ (\ref{eq:curious}). Then
\begin{align}
I_\epsilon(\br)
=\Re \frac{\ii}{r}\frac{\partial}{\partial\epsilon}\int_0^{2\pi}\frac{\dd\phi}{\cos\phi+\ii(\epsilon/r)}
=\Re \frac{\ii}{r}\frac{\partial}{\partial\epsilon}\int_0^{2\pi}\dd\phi\frac{\cos\phi-\ii(\epsilon/r)}{\cos^2\phi+(\epsilon/r)^2}.
\end{align}
The real part of the integral vanishes by symmetry, so that
\begin{align}
I_\epsilon(\br)=4\frac{\partial}{\partial\epsilon}\frac{\epsilon}{r^2}\int_0^{\pi/2}\frac{\dd\phi}{\cos^2\phi+(\epsilon/r)^2}
=2\pi\frac{\partial}{\partial\epsilon}\frac{1}{(r^2+\epsilon^2)^{1/2}}=-(2\pi)^2 \frac{\epsilon}{2\pi(r^2+\epsilon^2)^{3/2}}.
\end{align}
The fraction in this result goes to $\delta(\br)$ as $\epsilon\to 0$ \cite{KANW04}, which proves identity (\ref{eq:curious}).

\section{Angular integrals}
\label{sec:appflat}
We explain the computation of the $\bbm$ component of integral (\ref{eq:intflat1}). The other angular integrals of the paper are obtained by the same method. We thus need to compute
\begin{align}
\label{eq:ivmint}
I_m&
=\int_0^{2\pi}\frac{\dd\phi}{2\pi}\frac{p_\psi(\phi)\sin(\gamma-\phi)(-\sin\phi)}{\sin^2(\gamma-\phi)+\varepsilon^2\sin^2\phi}
\end{align}
where $\varepsilon>0$ stands for $a/r$, $p_\psi(\phi)=\tan(\psi-\phi)$, and $\psi$ is a complex angle. Then, the imaginary parts $\Im\psi$ and $\Im p$ have same signs. Introducing the complex variable $z=\exp(\ii\phi)$, integral (\ref{eq:ivmint}) is transformed into a contour integral over $z$ in the complex plane, on the unit circle $\Gamma$ defined by $|z|=1$. This change of variables entails
\begin{subequations}
\begin{align}
\label{eq:sumres}
I_m&=\int_\Gamma\frac{\dd z}{2\ii\pi}f_m(z)=\ii\sum {\rm Res}\{f_m(z)\}=\ii\sum_k w_k R_k,\\ f_m(z)&=\frac{\ee^{\ii\gamma}(z^2-1)(z^2-\ee^{2\ii\gamma})(z^2-\ee^{2\ii\psi})}{z(z^2+\ee^{2\ii\psi})
[(z^2-\ee^{2\ii\gamma})^2+\ee^{2\ii\gamma}\varepsilon^2(z^2-1)^2]},
\end{align}
\end{subequations}
where, invoking Cauchy's theorem, the integral is computed by summing up residues $R_k$ of $f_m$ at its poles. The first sum in (\ref{eq:sumres}) is over the residues at those poles $z_k$ of $f_m$ that are enclosed within the contour $\Gamma$ (i.e., such that $|z_k|<1$). The rightmost one, in which we have introduced weights $w_k$, is over the residues at \emph{all} poles, provided that we take $w_k=1$ if $|z_k|<1$ and $w_k=0$ if $|z_k|>1$. In the intermediate case $|z_k|=1$ where the pole lies \emph{on} the contour $\Gamma$, the principal-value prescription is invoked to handle the singularity. As discussed shortly, this simply amounts to taking $w_k=1/2$. The function $f_m(z)$ has seven poles $z_k$, $k=0,\ldots,6$ with associated residues $R_k$. With $s_y=\sign(\sin\gamma)=\sign y$, they read
\begin{align}
\begin{array}{ll}
\dps{z_0=0}&\dps{ R_0
=-\frac{\ee^{\ii\gamma}}{\ee^{2\ii\gamma}+\varepsilon^2}},\\
\dps{z_{1,2}=\pm\ii\ee^{\ii\psi}}&
\dps{R_{1,2}
=\frac{\cos(\gamma-\psi)\cos\psi}{\cos^2(\gamma-\psi)+\varepsilon^2\cos^2\psi}},\\
\dps{z_{3,4}=\pm\sqrt{\frac{1+\varepsilon^2-2\varepsilon|\sin\gamma|}{\ee^{-2\ii\gamma}+\varepsilon^2}}} & \dps{R_{3,4}=\frac{\sin\gamma[\sin(\gamma-\psi)+\ii s_y\varepsilon\sin\psi]}{2(\varepsilon^2-1+2\ii\varepsilon s_y \cos\gamma)[\cos(\gamma-\psi)-\ii s_y\varepsilon\cos\psi]}},\\
\dps{z_{5,6}=\pm\sqrt{\frac{1+\varepsilon^2+2\varepsilon|\sin\gamma|}{\ee^{-2\ii\gamma}+\varepsilon^2}}} &
\dps{R_{5,6}=\frac{\sin\gamma[\sin(\gamma-\psi)-\ii s_y \varepsilon\sin\psi]}{2(\varepsilon^2-1-2\ii\varepsilon s_y \cos\gamma)[\cos(\gamma-\psi)+\ii\varepsilon s_y\cos\psi]}}.
\end{array}
\end{align}
In the expressions for the poles $z_{3,4,5,6}$, we have for convenience correlated with $s_y$ the alternate sign under the square roots by introducing the absolute value of $\sin\gamma$, so that $|z_{3,4}|\leq 1$ (poles inside the contour) and $|z_{5,6}|\geq 1$ (poles outside the contour).  As a result, $\varepsilon$ always intervenes within a group $(s_y\varepsilon)$ in the residues. The latter inequalities are saturated if and only if $\gamma=0$ or $\gamma=\pi$, namely, on the the glide path $y=0$ in Cartesian coordinates. In the generic case $\gamma\not=0,\pi$, we have $w_{3,4}=1$ and $w_{5,6}=0$. When $\gamma\to 0,\pi$ the poles in the pairs $z_3$ and $z_5$, and $z_4$ and $z_6$, coalesce, pinching the contour: the integral develops a so-called \emph{pinch singularity}, which sometimes leads to contour integrals being divergent. However, the singularity is inessential in the present case since the residues $R_{3,4,5,6}$ vanish in the limit. Moreover, $w_0=1$, and  if $\Im\psi>0$, then  $|z_{1,2}|<1$ so that $w_{1,2}=1$, whereas $w_{1,2}=0$ in the opposite situation. It is intuitively obvious that the intermediate singular case $\Im\psi=0$ (for which both poles $z_{1,2}$ lie on the contour, and the contour integral is ill-defined) must be handled by taking $w_{1,2}=1/2$, i.e., by averaging the results of both previous cases. This amounts to taking a principal-value prescription.

Expressing the weighted sum of residues (\ref{eq:sumres}) in terms of $p^0=\tan\psi$ by means of lengthy simplifications involving usual trigonometric identities, we compute separately the cases $\Im\psi>0$ (only poles $z_{0,3,4}$ contribute) and $\Im\psi<0$ (only poles $z_{0,1,2,3,4}$ contribute). Both results are encoded into the following single expression, where $s_\alpha=\sign\Im\psi=\sign\Im p^0$ with the convention that $\sign(0)=0$:
\begin{align}
I_m
&=-\frac{s_y\varepsilon+\sin\gamma}{1+\varepsilon^2+2\varepsilon|\sin\gamma|}
+\frac{s_y\varepsilon+\ii s_\alpha(\cos\gamma+p^0\sin\gamma)}{\varepsilon^2+(\cos\gamma+p^0\sin\gamma)^2}\nonumber\\
&=-s_y\frac{|\sin\gamma|+\varepsilon}{\cos^2\gamma+(\varepsilon+|\sin\gamma|)^2}
-\frac{1}{\ii}\frac{1}{s_\alpha(\cos\gamma+p^0\sin\gamma)+\ii s_y\varepsilon}\nonumber\\
&=r\left[-s_y\frac{|y|+a}{x^{\prime2}+(|y|+a)^2}
-\frac{1}{\ii}\frac{1}{s_\alpha(x'+p^0 y)+\ii s_y a}\right].
\end{align}
Multiplying by $1/r$ as in Eq.\ (\ref{eq:intflat1}) eventually yields the $\bbm$ component in that equation. Calculations as above are eased by the use of an algebraic computational toolbox.


\end{document}